\documentclass[%
 reprint,showkeys,
 amsmath,amssymb,
 aps,
]{revtex4-2}

\usepackage{graphicx}
\usepackage{dcolumn}
\usepackage[usenames]{color}
\usepackage{colortbl}
\usepackage{bm}
\usepackage{longtable}

\begin{document}

\title{Cross-section of the ${^{95\rm}\rm{Nb}}$ production on natural molybdenum 
\\	at the bremsstrahlung end-point energy up to 95 MeV}


\author{I.S. Timchenko$^{1,2,}$} \email{iryna.timchenko@savba.sk;\\ timchenko@kipt.kharkov.ua}
\author{O.S. Deiev$^2$, S.M. Olejnik$^2$, S.M. Potin$^2$, \\
	V.A. Kushnir$^2$, V.V. Mytrochenko$^{2,3}$, S.A. Perezhogin$^2$}

\affiliation{$^1$ Institute of Physics, Slovak Academy of Sciences, SK-84511 Bratislava, Slovakia}%
\affiliation{$^2$ National Science Center "Kharkov Institute of Physics and Technology", 1, Akademichna St., 61108, Kharkiv, Ukraine}%
\affiliation{$^3$ CNRS/IJCLAB, 15 Rue Georges Clemenceau Str., Orsay, 91400, France}%

\date{\today}

\begin{abstract}
The photoproduction of $^{95\rm m}$Nb on ${^{\rm nat}\rm{Mo}}$ was studied using the electron beam of the LUE-40 linac RDC "Accelerator" NSC KIPT. Measurements were performed using activation and off-line $\gamma$-ray spectrometric techniques. The experimental flux-averaged cross-section $\langle{\sigma(E_{\rm{\gamma max}})}\rangle_{\rm{m}}$ for the ${^{\rm nat}\rm{Mo}}(\gamma,x\rm np)^{95\rm m}$Nb reaction at the bremsstrahlung end-point energy range of 38--93 MeV has been first time obtained. The estimated values $\langle{\sigma(E_{\rm{\gamma max}})}\rangle_{\rm{g}}$ for the formation of $^{95}$Nb in the ground state and total cross-sections $\langle{\sigma(E_{\rm{\gamma max}})}\rangle_{\rm{tot}}$ for the studied reaction were determined. 
The theoretical values of the yields $Y_{\rm m,g,tot}(E_{\rm{\gamma max}})$  and flux-averaged cross-sections $\langle{\sigma(E_{\rm{\gamma max}})}\rangle_{\rm{m,g,tot}}$ for the ${^{\rm nat}\rm{Mo}}(\gamma,x\rm np)^{95\rm m,g,tot}$Nb reactions were calculated using the cross-sections $\sigma(E)$ from the TALYS1.95 code for six different level density models. The comparison showed a significant excess of the experimental results over the theoretical  $\langle{\sigma(E_{\rm{\gamma max}})}\rangle_{\rm{m,g,tot}}$ values. 
\end{abstract}

\maketitle

\section{Introduction}
\label{intro}

Experimental studies of photonuclear reactions are of particular interest today. This is due to the fact that there is a problem of discrepancy between experimental data on cross-sections $(\gamma,\rm n)$, $(\gamma,2\rm n)$, $(\gamma,3\rm n)$, measured in different laboratories \cite{1,2,Wol1,Wol2,Varl1,Varl2,Varl3,Varl4,Varl5}. The development of modern theoretical models to describe the mechanisms of nuclear reactions and program codes \cite{empire,talys,ccone,coh3} based on these concepts requires verification using data on multiparticle photonuclear reactions $(\gamma,x\rm n)$ and $(\gamma,x{\rm n}y{\rm p})$. In addition, there is a lack of such data in international databases \cite{exfor,endf,ripl,cdfe,LUND}.

Multiparticle photonuclear reactions have small cross-sections that can be measured using intense fluxes of incident $\gamma$-quanta \cite{Meth2,arxivHf,arxivNb,Al}. Such $\gamma$-quanta fluxes can be produced by linear electron accelerators using a converter to generate bremsstrahlung radiation. This type of experiment allows measuring integral characteristics of reactions, such as reaction yield \cite{IR8,CS3}, flux-average  cross-section \cite{Meth4,Cu,IR6,IR7}, cross-section per equivalent photon \cite{Ma78,Ma00,DiNap}, etc., and requires additional mathematical processing of the results. Despite this, bremsstrahlung remains an important tool in the study of photonuclear reactions.

Experiments on the photodisintegration of stable isotopes of the Mo nucleus were described in literature, for example, \cite{IR8,CS3,arxivNb,Ma78,15,16,17,18,19,20,21,CS1,CS2}. 
However, the photodisintegration of the Mo nucleus with the formation of  $^{95}\rm{Nb}$  has not been studied much. The values of the isomer ratio for the $^{95\rm m,g}\rm{Nb}$ nuclei-product in reactions on natural molybdenum were mainly obtained \cite{15,16,17,18,19}. The data obtained in the region of Giant Dipole Resonance have been presented in different ways and some of them contradict each other \cite{18,19}. 

At an intermediate energy region for the isomer ratios for the $^{\rm nat}$Mo$(\gamma,x\rm np)^{95\rm m,g}$Nb  reaction were measured in Ref.~\cite{IR8,CS3,arxivNb,20,21}. All these data are in good agreement with each other, but show a marked difference from the theoretical predictions based on cross-sections $\sigma(E)$ from the TALYS1.95 code for six different level density models (see Ref. ~\cite{arxivNb}). For a more detailed analysis of theoretical models, it is necessary to compare theoretical estimates with experimental cross-sections of the ${^{\rm nat}\rm{Mo}}(\gamma,x\rm np)^{95\rm m,g}$Nb reactions. 

There is little data in the literature on a cross-section of the reaction ${^{\rm nat}\rm{Mo}}(\gamma,x\rm np)^{95\rm m}$Nb. The activation yield curves have been presented in \cite{Ma78} for a number of photonuclear reactions in the energy range from 30 to 68~MeV, in order to evaluate quantitatively the interferences due to competing reactions in multielement photon activation analysis. As a result, in \cite{Ma78} the energy dependence of the yield of the ${^{96}\rm{Mo}}(\gamma,x\rm np)^{95\rm m}$Nb reaction was obtained in the terms of the average cross-section per equivalent photon.

In this work a study of the $^{95\rm m,g}$Nb production in photonuclear reactions on ${^{\rm nat}\rm{Mo}}$ at the bremsstrahlung end-point energy range of $E_{\rm{\gamma max}}$ = 38--93~MeV is performed. The obtained experimental results are compared with data from Ref.~\cite{Ma78} and theoretical estimates performed with the cross-sections $\sigma(E)$ from the TALYS1.95 code with different level density models $LD$~1-6. 

\section{Experimental setup and procedure}
\label{sec:1}

\begin{figure}[h]
	\resizebox{0.49\textwidth}{!}{%
		\includegraphics{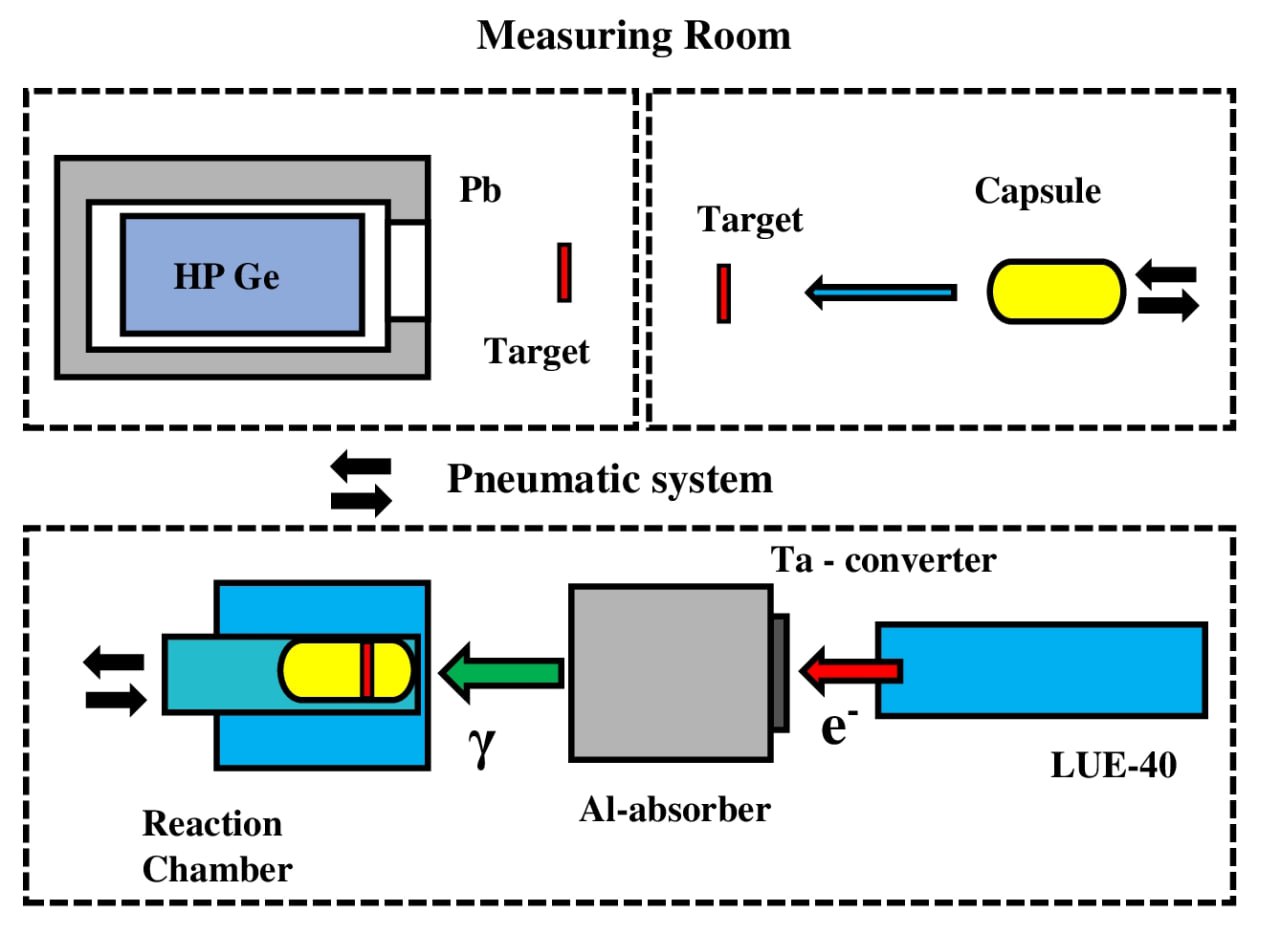}}
	\caption{Block diagram of the experiment. The lower part shows the linear accelerator LUE-40, the Ta converter, the Al absorber, and the exposure reaction chamber. The upper part shows the measuring room. }
	\label{fig1}
\end{figure}

The experimental measuring complex of our laboratory for the study of photonuclear reactions and in particular for the study of the formation of the $^{95\rm m}$Nb nucleus in reactions on $^{\rm nat}$Mo is shown in the form of a block diagram in Fig.~\ref{fig1}. 

The experiment was performed using the electron linear accelerator LUE-40 of Research and Development Center "Accelerator" of the National Science Center "Kharkov Institute of Physics and Technology" at the range of initial electron energy is $E_{\rm e}$ = 30--100~MeV. Previously, we described in detail the parameters of the linear accelerator LUE-40, which can be found in \cite{Li1,Li2,Li3,Li4}. Description of the details of the experiment (irradiation of targets, measurement of spectra of $\gamma$-induced activity,  modeling of the bremsstrahlung fluxes using the  GEANT4.9.2 code\cite{GE}, and procedure of monitoring the bremsstrahlung flux with the yield of the $^{100}\rm{Mo}(\gamma,n)^{99}\rm{Mo}$ reaction) can be found in papers \cite{Meth1,Meth2,Meth3,Meth4,Nb95xn}.

For the experiment the targets were made of natural molybdenum. The shape of the targets were thin foil discs with a diameter of 8~mm and a thickness of $\approx$0.11~mm,which corresponds to a mass of $\approx$60~mg. Natural molybdenum consists of 7 stable isotopes, with isotopic abundances: $^{92}$Mo -- 0.1484, $^{94}$Mo -- 0.0925, $^{95}$Mo -- 0.1592, $^{96}$Mo -- 0.1668, $^{97}$Mo -- 0.0955, $^{98}$Mo -- 0.2413, $^{100}$Mo -- 0.0963 (according \cite{talys,GE}).

\begin{figure*}[]
	\begin{minipage}[h]{0.99\linewidth}
		{\includegraphics[width=1\linewidth]{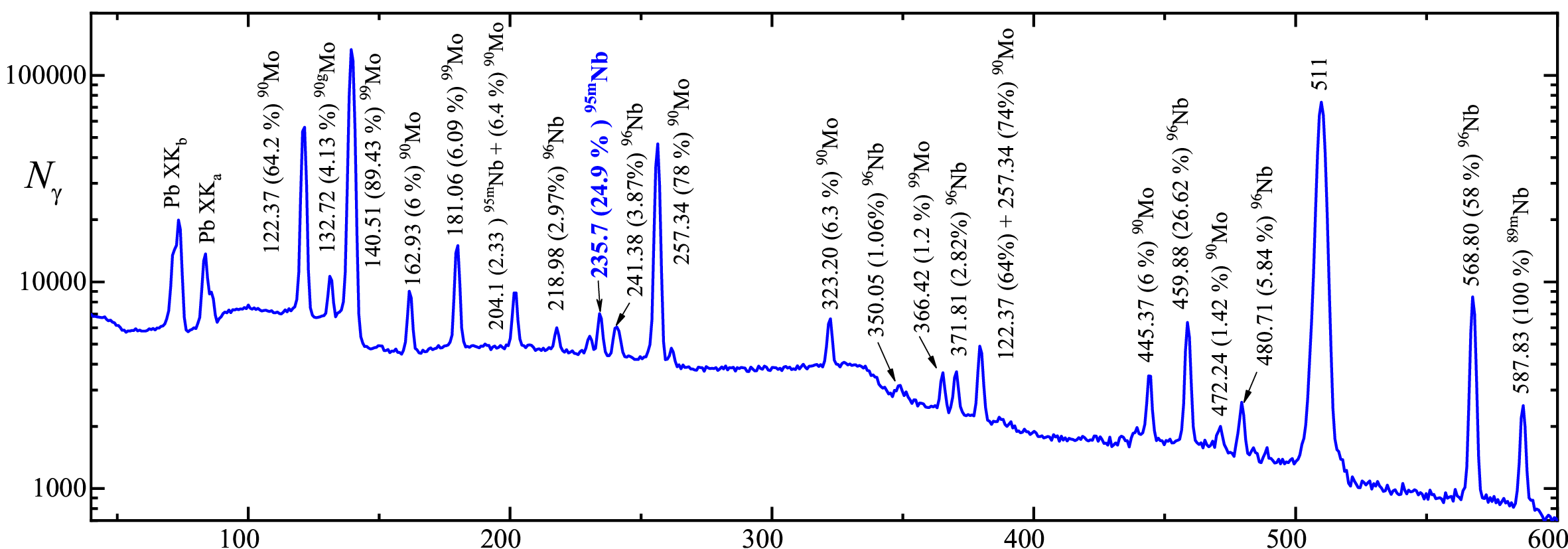}} \\
	\end{minipage}
	\vfill
	\begin{minipage}[h]{0.99\linewidth}
		{\includegraphics[width=1\linewidth]{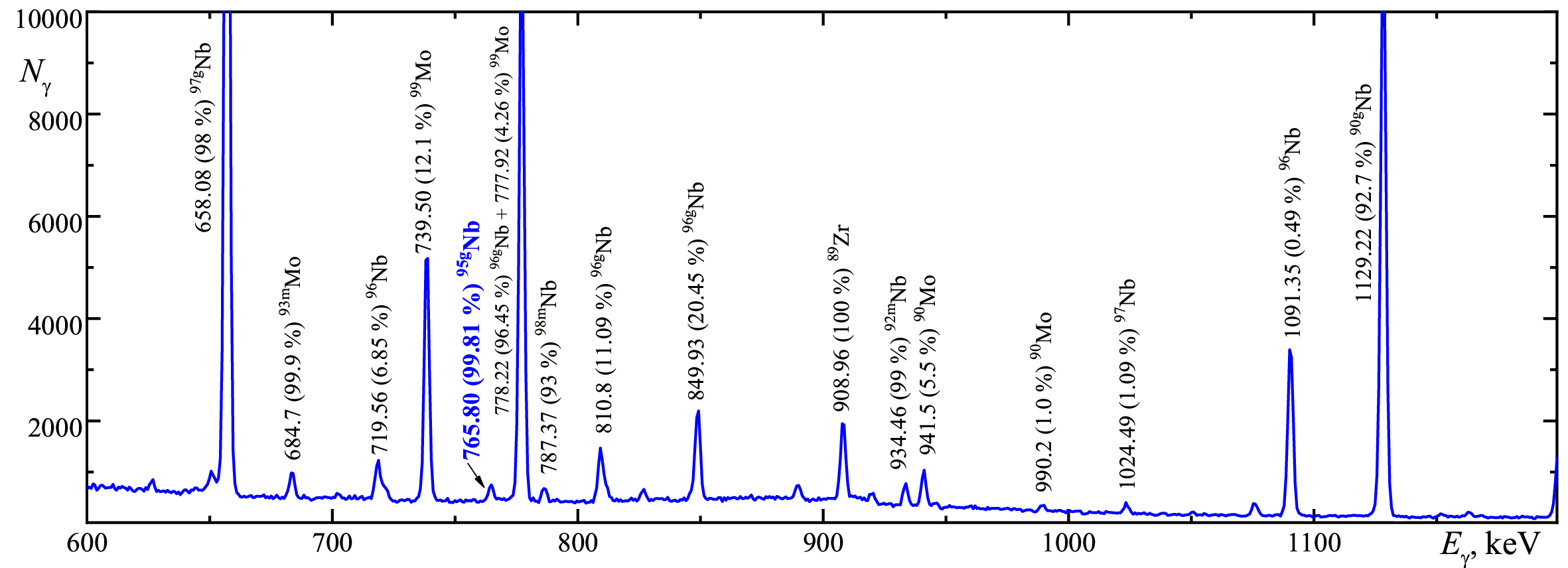}} \\
	\end{minipage}
	\caption{Two fragments of $\gamma$-ray spectrum in the energy ranges $40 \leq E_{\rm \gamma} \leq 600$~keV and $600 \leq E_{\rm \gamma} \leq 1200$~keV from the $^{\rm nat}$Mo target of mass 57.862~mg after irradiation of the bremsstrahlung $\gamma$-flux at the end-point energy $E_{\rm{\gamma max}}$ = 92.50~MeV. The irradiation $t_{\rm irr}$ and measurement $t_{\rm meas}$ times were both 3600~s.}
	\label{fig2}
\end{figure*}

As an example, the $\gamma$-radiation spectrum of a $^{\rm nat}$Mo target is shown in Fig.~\ref{fig2}. As can be seen, the $\gamma$-radiation spectrum of a natural molybdenum target is a complex pattern. There are many peaks corresponding the $\gamma$-lines from the $^{\rm nat}$Mo$(\gamma,x{\rm n}y{\rm p})$ reactions.

\section{Calculation of cross-sections, flux-averaged cross-sections, and yields of photonuclear reactions  }
\label{sec:2}
The photonuclear reactions $^{\rm nat}$Mo$(\gamma,x{\rm np})$ nucleus formed metastable and ground states of $^{95\rm m,g}$Nb. Natural molybdenum consists of seven stable isotopes, but
only four isotopes took part in the formation of the $^{95}$Nb
nucleus. Respectively, there are four reactions with different thresholds: 

$^{96}$Mo$(\gamma,\rm p)^{95\rm g}$Nb -- $E_{\rm thr}$ = 9.30 MeV;

$^{97}$Mo$(\gamma,\rm np)^{95\rm g}$Nb -- $E_{\rm thr}$ = 16.12 MeV;

$^{98}$Mo$(\gamma,\rm 2np)^{95\rm g}$Nb -- $E_{\rm thr}$ = 24.76 MeV;

$^{100}$Mo$(\gamma,\rm 4np)^{95\rm g}$Nb -- $E_{\rm thr}$ = 38.98 MeV. 

The thresholds for the metastable state $^{95\rm m}$Nb formation nucleus are higher than in the ground state with an additional excitation energy of 235.7 keV. 

The theoretical cross-sections $\sigma(E)$ of studied reactions for monochromatic photons were calculated using the TALYS1.95 open code \cite{talys} for different level density models $LD$ 1--6. In the TALYS code, there are three phenomenological level density models and three options for microscopic level densities. Descriptions of the models are given in \cite{talys}:

$LD 1$: Constant temperature + Fermi gas model, introduced by Gilbert and Cameron \cite{LD1}.

$LD 2$: Back-shifted Fermi gas model \cite{LD2}. 

$LD 3$: Generalized superfluid model (GSM) \cite{LD31,LD32}. 

$LD 4$: Microscopic level densities (Skyrme force) from Goriely’s tables \cite{LD4}. 

$LD 5$: Microscopic level densities (Skyrme force) from Hilaire’s combinatorial tables \cite{LD5}. 

$LD 6$: Microscopic level densities based on temperature-dependent Hartree-Fock-Bogoliubov calculations using the Gogny force \cite{LD6} from Hilaire’s combinatorial tables.

Figs.~\ref{fig3} (a--c) show the calculated cross-sections $\sigma(E)$ for the formation of the $^{95\rm m,g,tot}$Nb nucleus on 4 stable isotopes of molybdenum: $^{96}$Mo, $^{97}$Mo, $^{98}$Mo, and $^{100}$Mo. These cross-sections take into account the percentage abundance of isotopes. In all three cases, the cross-sections for natural molybdenum (black curve) were obtained as the sum of the cross-sections for 4 isotopes with the percentage abundance of isotopes.
	
As can be seen from Fig.~\ref{fig3} (a), at an energy range of the GDR, the main contribution to the formation of the nucleus $^{95\rm m}$Nb is given by the  $^{96}$Mo$(\gamma,\rm p)$ reaction. For energy above 30 MeV, all other isotopes need to be taken into account for the formation of the $^{95\rm m}$Nb. In the energy dependence of the cross-section of the $^{95\rm m}$Nb nucleus formation, the dominance of a contribution of the $^{96}$Mo isotope is noticeable, while for  $^{95\rm g}$Nb the contribution of the cross-sections on the four Mo isotopes is more evenly distributed. The calculations have been performed for the TALYS1.95 code, level density model $LD$1.

Note that the results calculated for other level density models retain a similar tendency between the contributions of 4 isotopes to the cross-sections for the formation of the $^{95}$Nb nucleus (dominance of a contribution of the $^{96}$Mo isotope for $^{95\rm m}$Nb, and more balanced contribution for $^{95\rm g,tot}$Nb).

The calculated reaction yield, using theoretical cross-sections $\sigma(E)$, is determined by the formula:

\begin{equation}\label{form1}
	Y(E_{\rm{\gamma max}}) = N_n \int\limits_{E_{\rm{thr}}}^{E_{\rm{\gamma max}}}\sigma(E)\cdot W(E,E_{\rm{\gamma max}})dE,
\end{equation}
where $N_n$ is the number of atoms of the element under study, $W(E,E_{\rm{\gamma max}})$  is the bremsstrahlung $\gamma$-flux. 

For the estimation of the contribution of a reaction in the total production of a studied nuclide (for example, the $^{96}$Mo$(\gamma,\rm p)$ reaction in the production of the $^{95}$Nb nucleus on $^{\rm nat}$Mo), the normalized reaction yield $Y_{\rm i}(E_{\rm{\gamma max}})$ was used. For calculation of $Y_{\rm i}(E_{\rm{\gamma max}})$ it was used the expression:

  \begin{figure}[h]
	\resizebox{0.45\textwidth}{!}{%
		\includegraphics{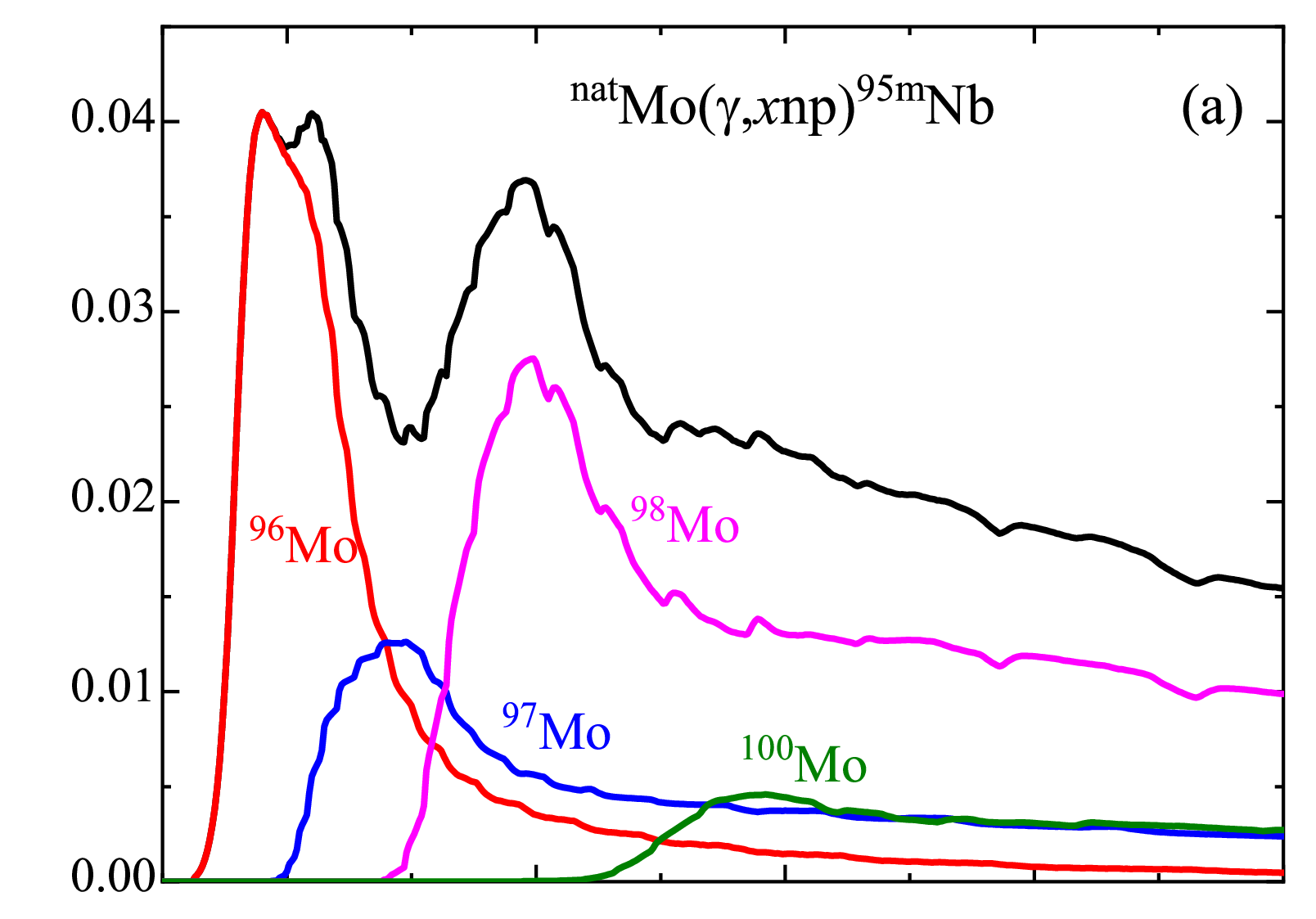}}
	\vspace{-1ex}
		\resizebox{0.45\textwidth}{!}{%
			\includegraphics{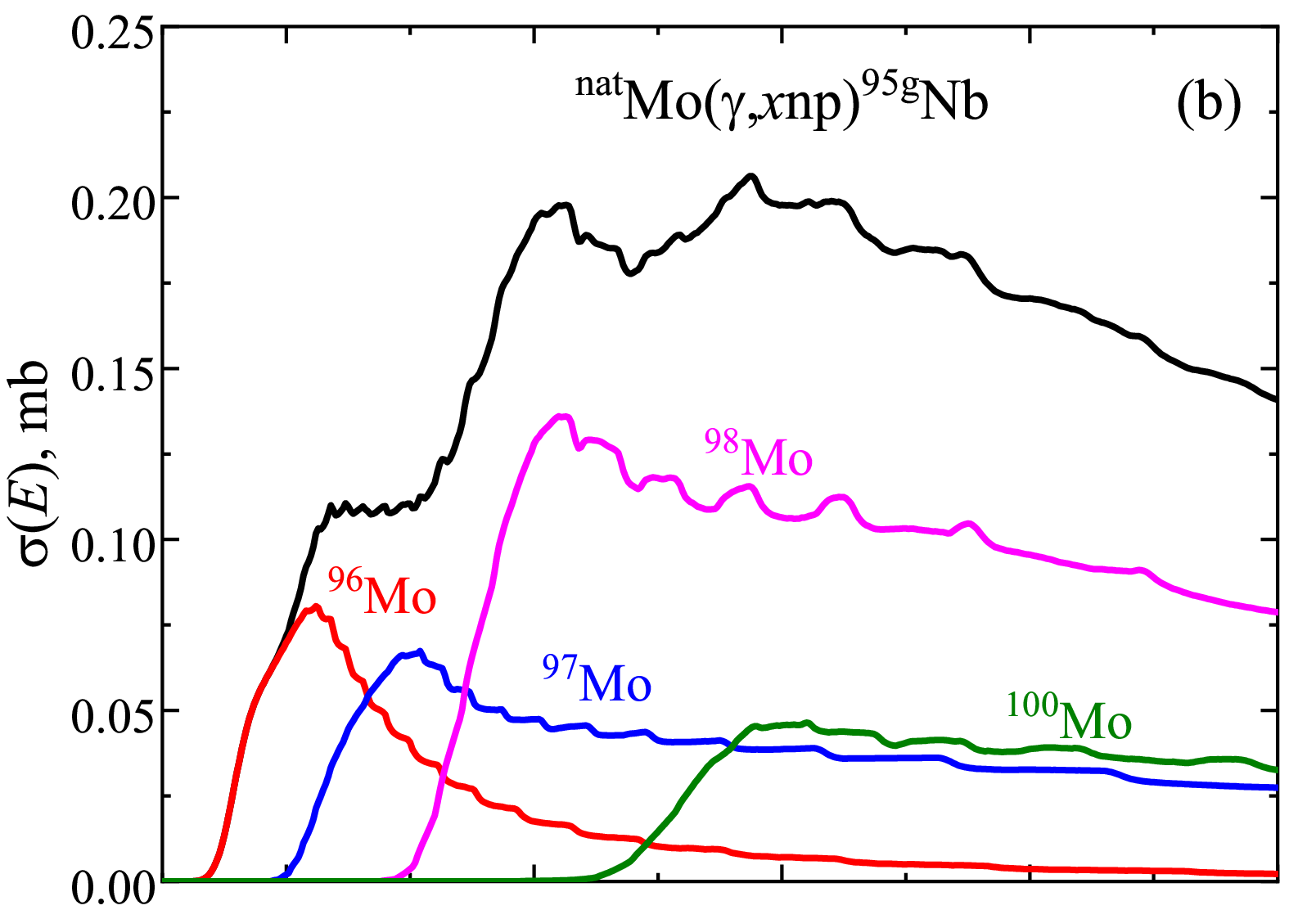}}
		\vspace{-1ex}
			\resizebox{0.45\textwidth}{!}{%
				\includegraphics{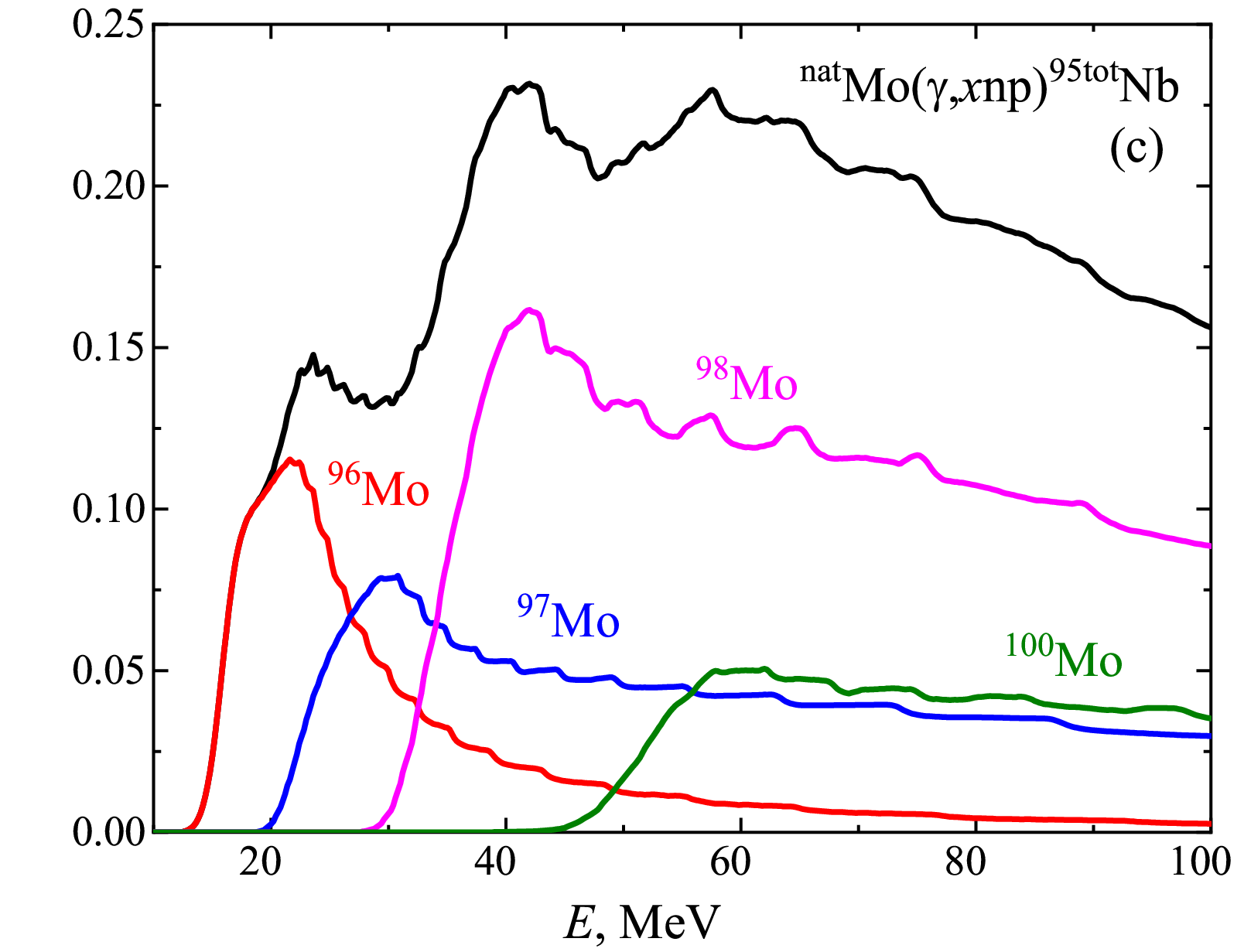}}
			\caption{Theoretical cross-sections $\sigma(E)$ for the formation of the $^{95\rm m,g,tot}$Nb nucleus on 4 stable isotopes of molybdenum (taking into account isotope percentage abundance) and on $^{\rm nat}$Mo (black curve). Results for production of the $^{95}$Nb nucleus in a metastable (a), and ground (b) states, total cross-sections (c) are shown. The calculations were performed in TALYS1.95 code, $LD$1.}
			\label{fig3}
		\end{figure}
	
  \begin{figure}[h]
	\resizebox{0.45\textwidth}{!}{%
		\includegraphics{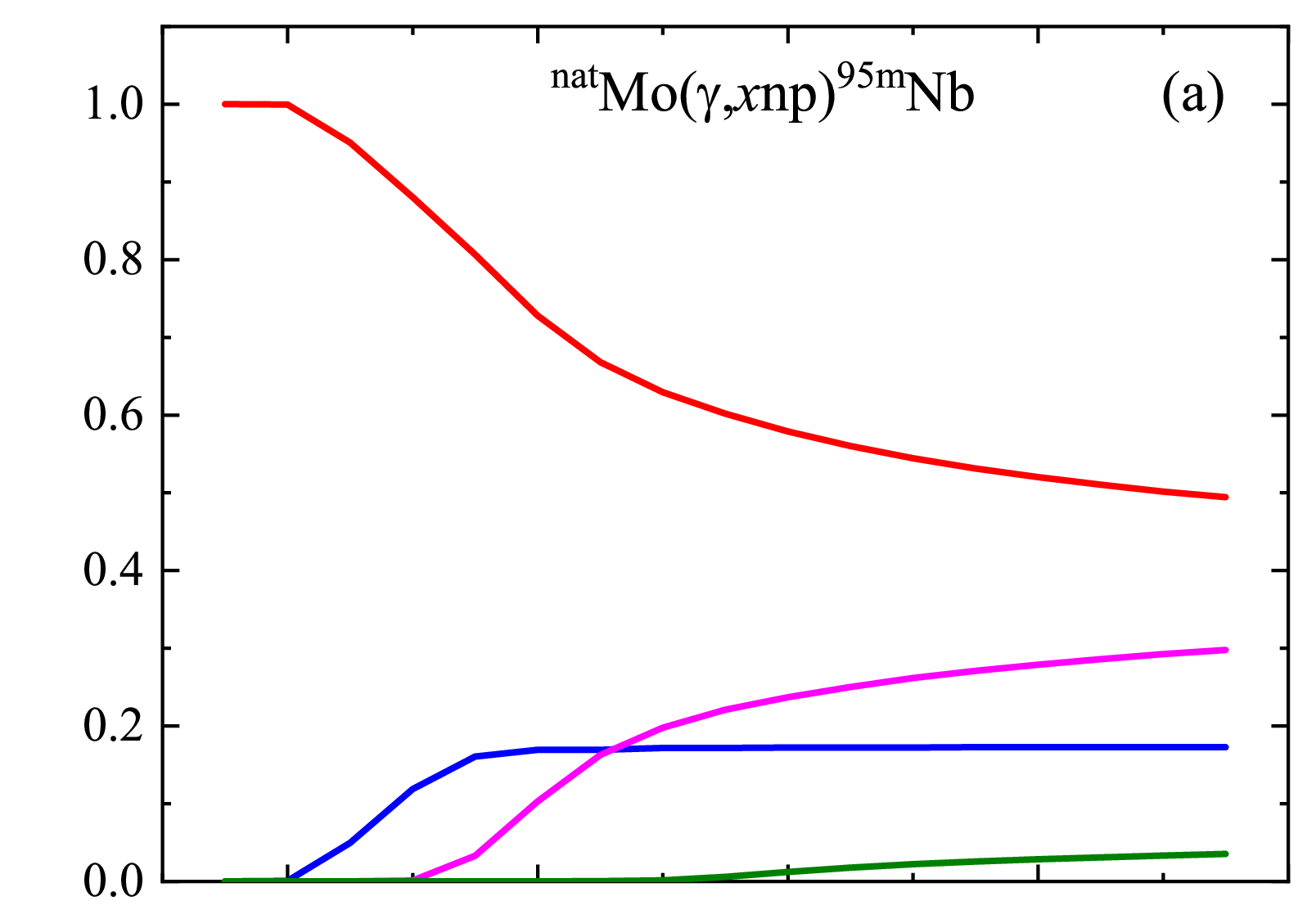}}
	\vspace{-1ex}
		\resizebox{0.45\textwidth}{!}{%
			\includegraphics{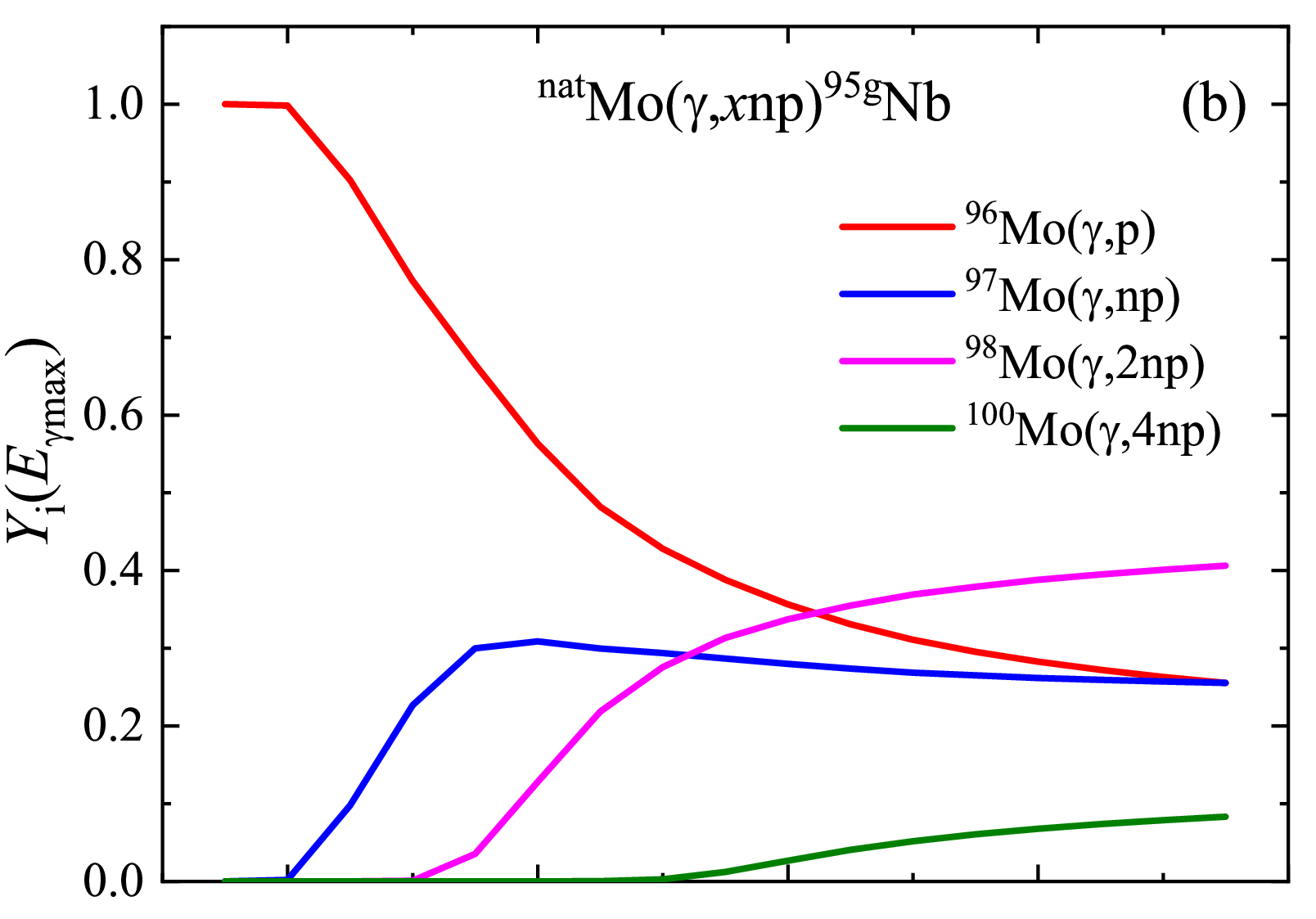}}
		\vspace{-1ex}
			\resizebox{0.45\textwidth}{!}{%
				\includegraphics{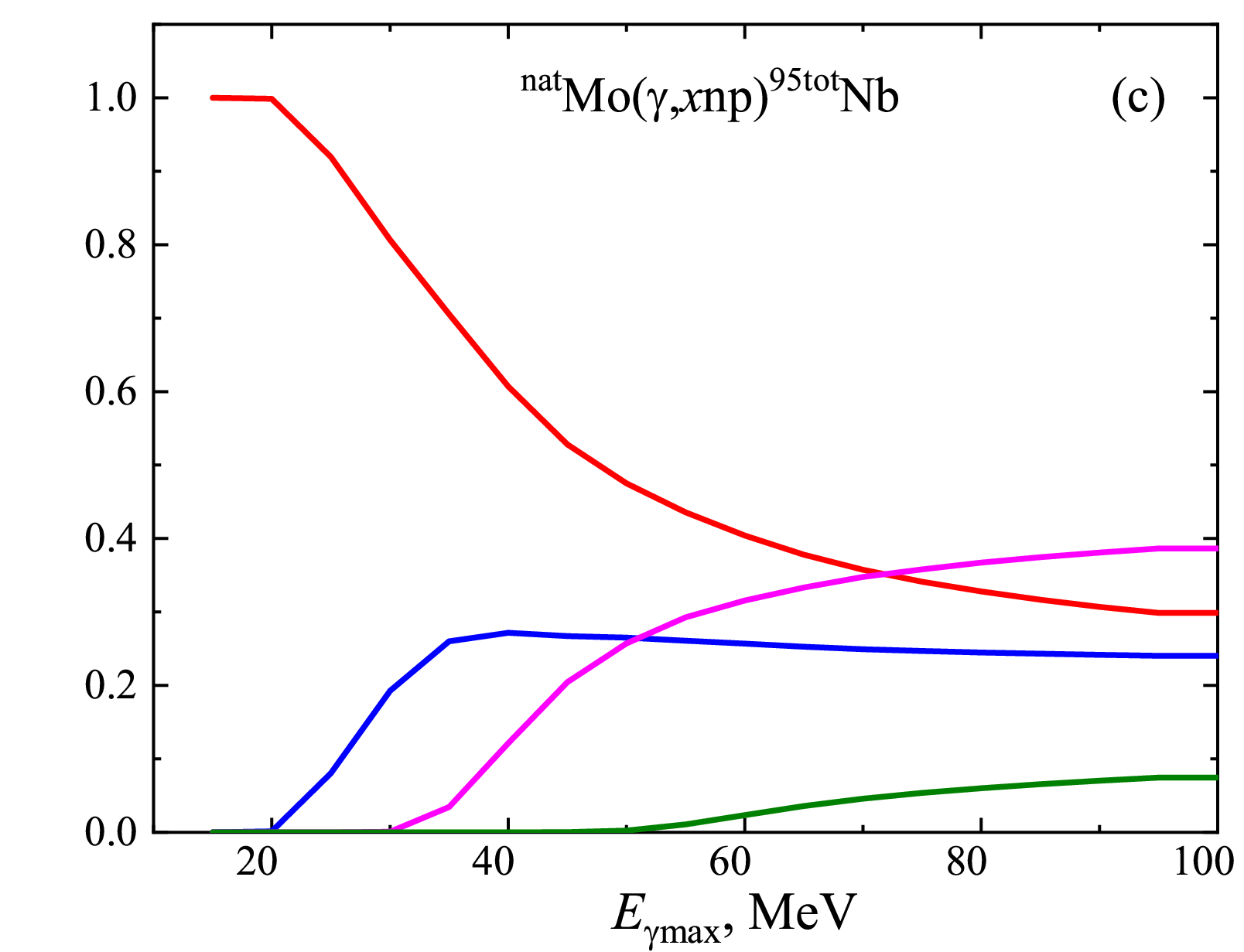}}
			\caption{The normalized reaction yields of the $^{95}$Nb formation on various molybdenum isotopes. The sum of the isotope normalized yields is equal to 1. Results for the formation of the $^{95}$Nb nucleus in the metastable (a), in the ground (b) states, and for total cross-section (c) are shown. The calculations were performed in the TALYS1.95 code, $LD$1.}
			\label{fig4}
		\end{figure}

     \begin{equation}\label{form2}
	Y_{\rm i}(E_{\rm{\gamma max}}) = \frac
	{A_i \int\limits_{E^i_{\rm{thr}}}^{E_{\rm{\gamma max}}}\sigma_i(E)\cdot W(E,E_{\rm{\gamma max}})dE}
	{\sum^4_{k=1} A_k \int\limits_{E^k_{\rm{thr}}}^{E_{\rm{\gamma max}}}\sigma_k(E)\cdot W(E,E_{\rm{\gamma max}})dE},
\end{equation}
where $\sigma_k(E)$ is the cross-section for the formation of the $^{95}$Nb nucleus on the $k$-th isotope with isotopic abundance $A_k$. Summation over $k$ was carried out for 4 stable molybdenum isotopes $^{96,97,98,100}$Mo. 

Figs.~\ref{fig4} (a--c)  show the normalized yields of various isotopes of the reaction $^{\rm nat}$Mo$(\gamma,x\rm np)^{95m,g,tot}$Nb according to Eq.~\ref{form2}. The value of the normalized yield on a given isotope is determined by the cross-section, the reaction threshold, and the isotope percentage abundance. 

As a rule, in the presence of several isotopes, there is one whose contribution to the reaction yield dominates ($>$ 90\%), as shown, for example, in Refs.~\cite{Cu,natNi-5756Ni}. In the case of the reaction $^{\rm nat}$Mo$(\gamma,x\rm np)^{95\rm m}$Nb at energies up to 20 MeV, the contribution of $^{96}$Mo is 100\%. With increasing energy, the contribution from isotope $^{96}$Mo decreases but remains dominant over the entire energy range under study. In the case of the formation of the $^{95}$Nb nucleus in the ground state at bremsstrahlung end-point energy above 30 MeV, it is difficult to determine the dominant reaction. Thus, for the estimation of the yield of the $^{\rm nat}$Mo$(\gamma,x\rm np)$ reaction with the production of the $^{95}$Nb nucleus in the ground state it is necessary to take into account the contribution of the four stable isotopes. 

\begin{figure}[h]
	\resizebox{0.45\textwidth}{!}{%
		\includegraphics{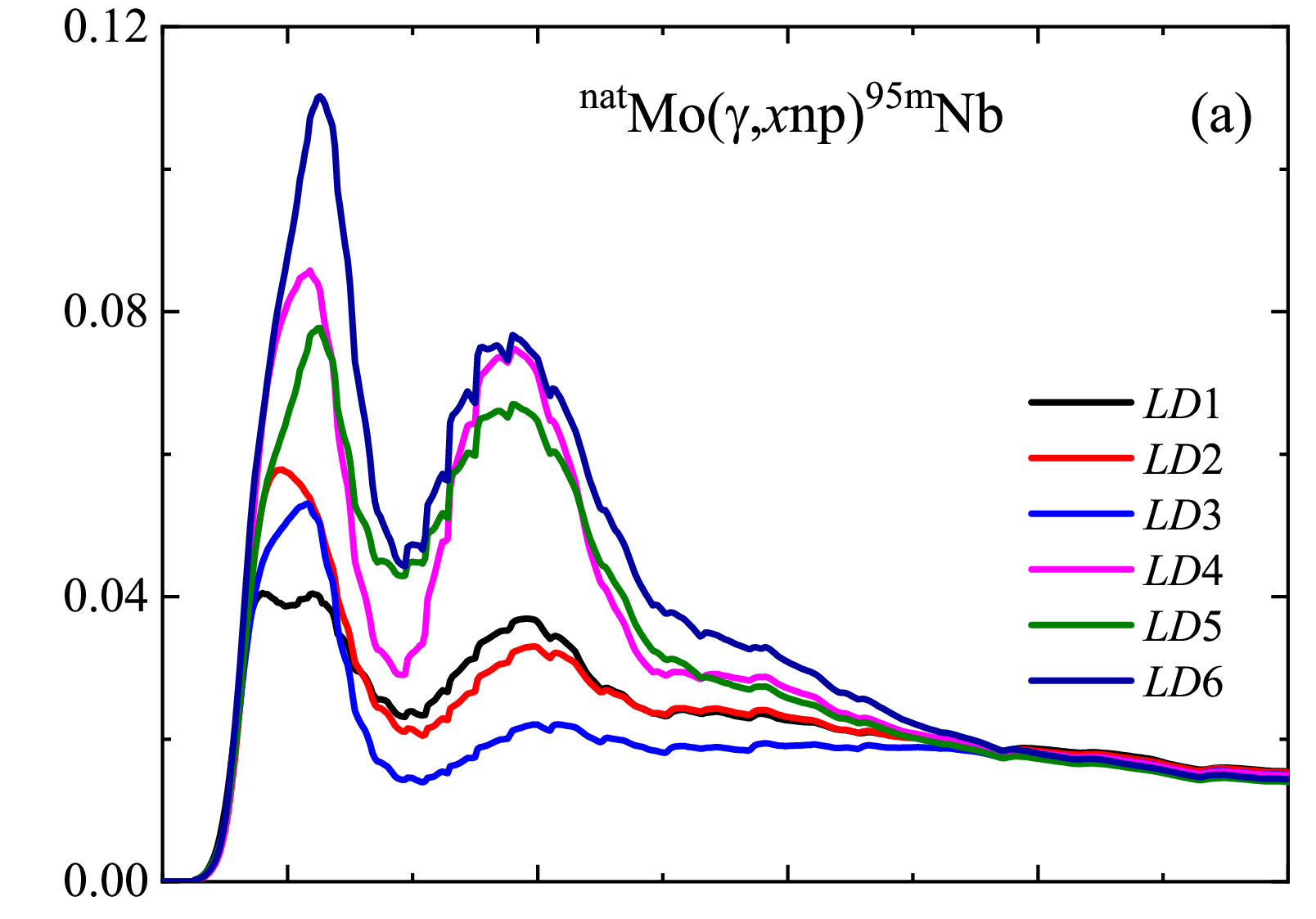}}
	\vspace{-1ex}
		\resizebox{0.45\textwidth}{!}{%
			\includegraphics{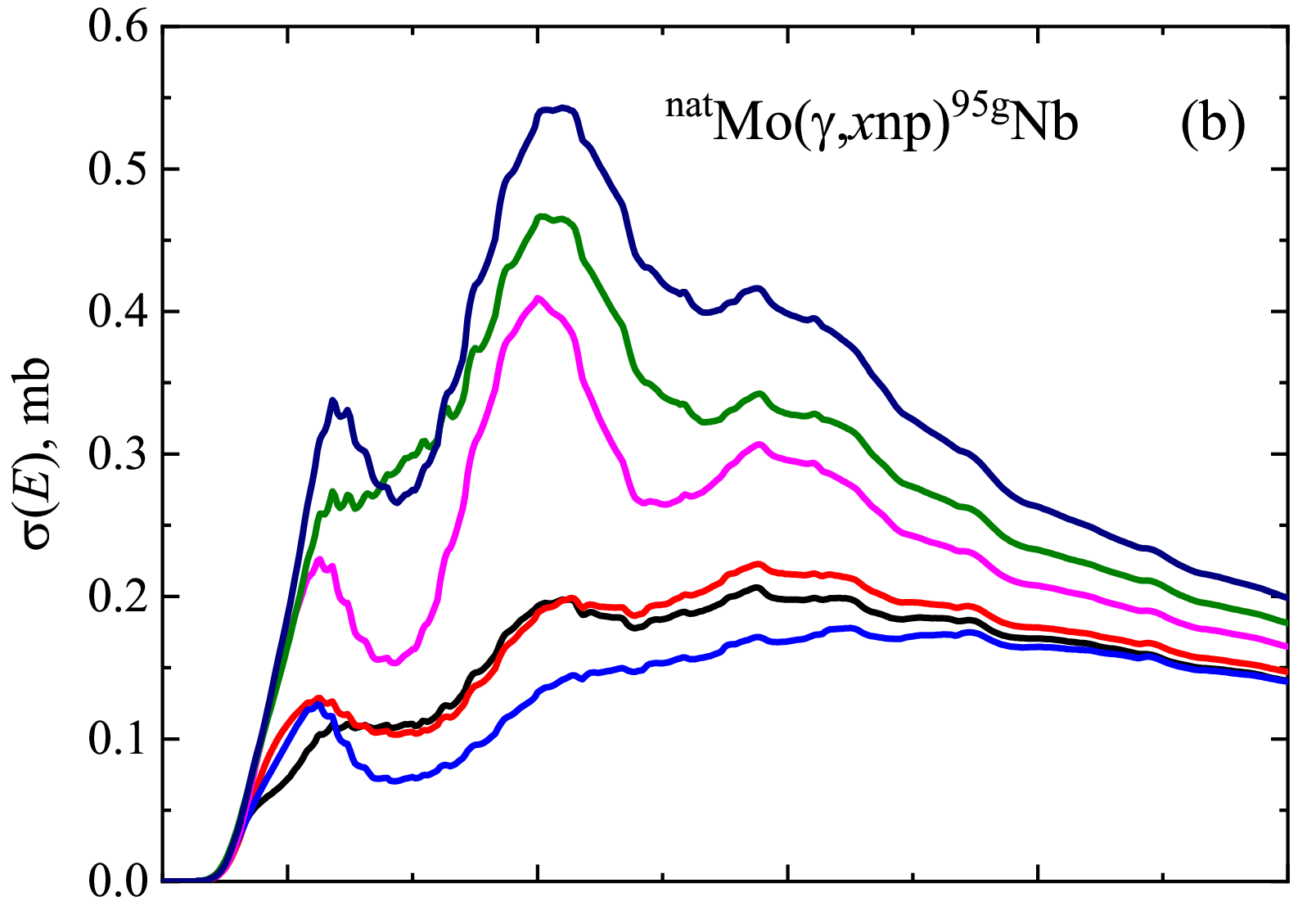}}
		\vspace{-1ex}
			\resizebox{0.45\textwidth}{!}{%
				\includegraphics{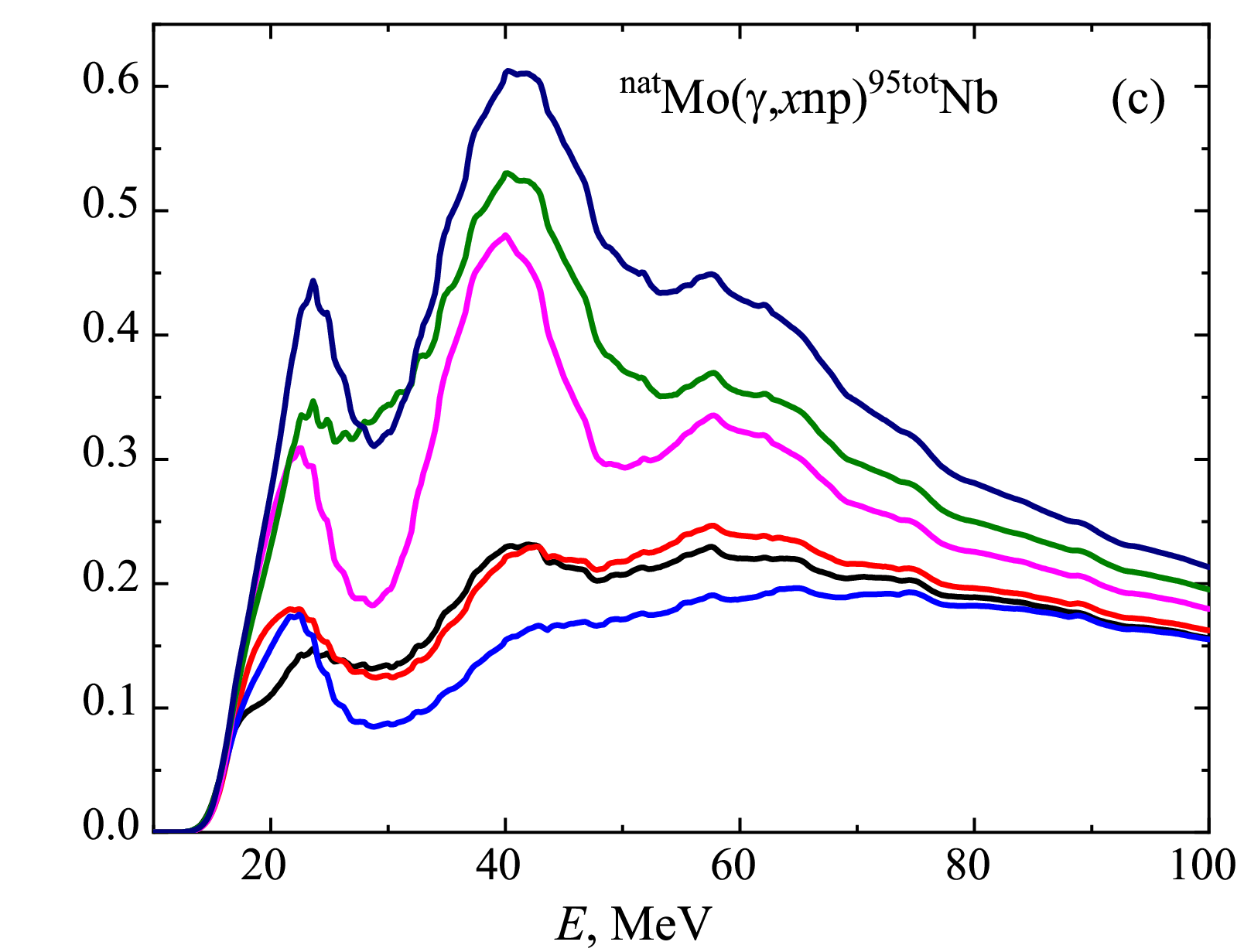}}
			\caption{Theoretical cross-sections $\sigma(E)$ of the $^{95\rm m,g,tot}$Nb formation on $^{\rm nat}$Mo, TALYS1.95 code for six different level density models. }
			\label{fig5}
		\end{figure}

The cross-sections $\sigma(E)$ for $^{\rm nat}$Mo$(\gamma,x\rm np)^{95\rm m,g,tot}$Nb reactions, calculated in TALYS1.95 for level density models $LD$ 1--6, are shown in Figs.~\ref{fig5} (a--c).

					\begin{figure}[h]
		\resizebox{0.45\textwidth}{!}{%
			\includegraphics{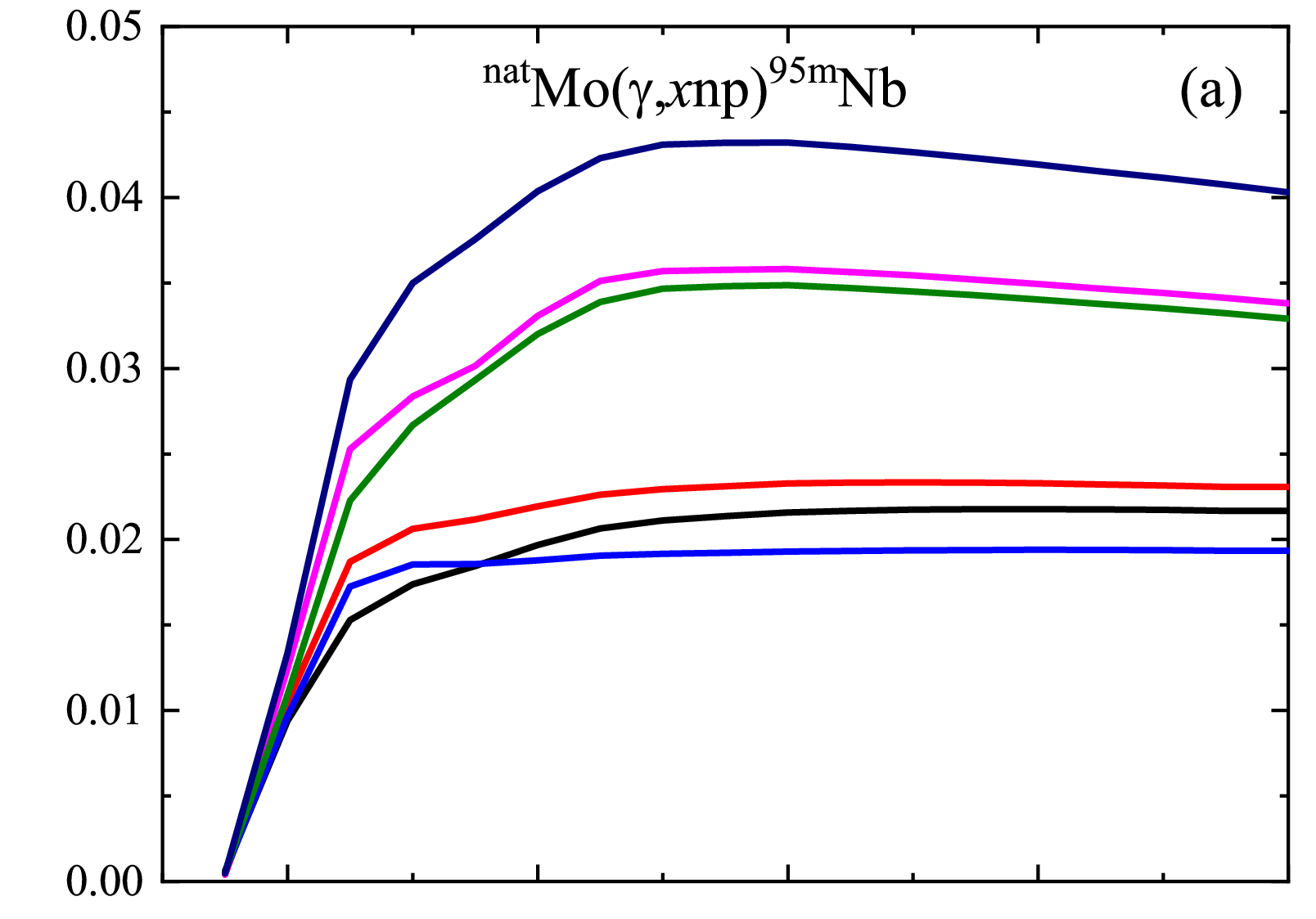}}
		\vspace{-1ex}
			\resizebox{0.45\textwidth}{!}{%
				\includegraphics{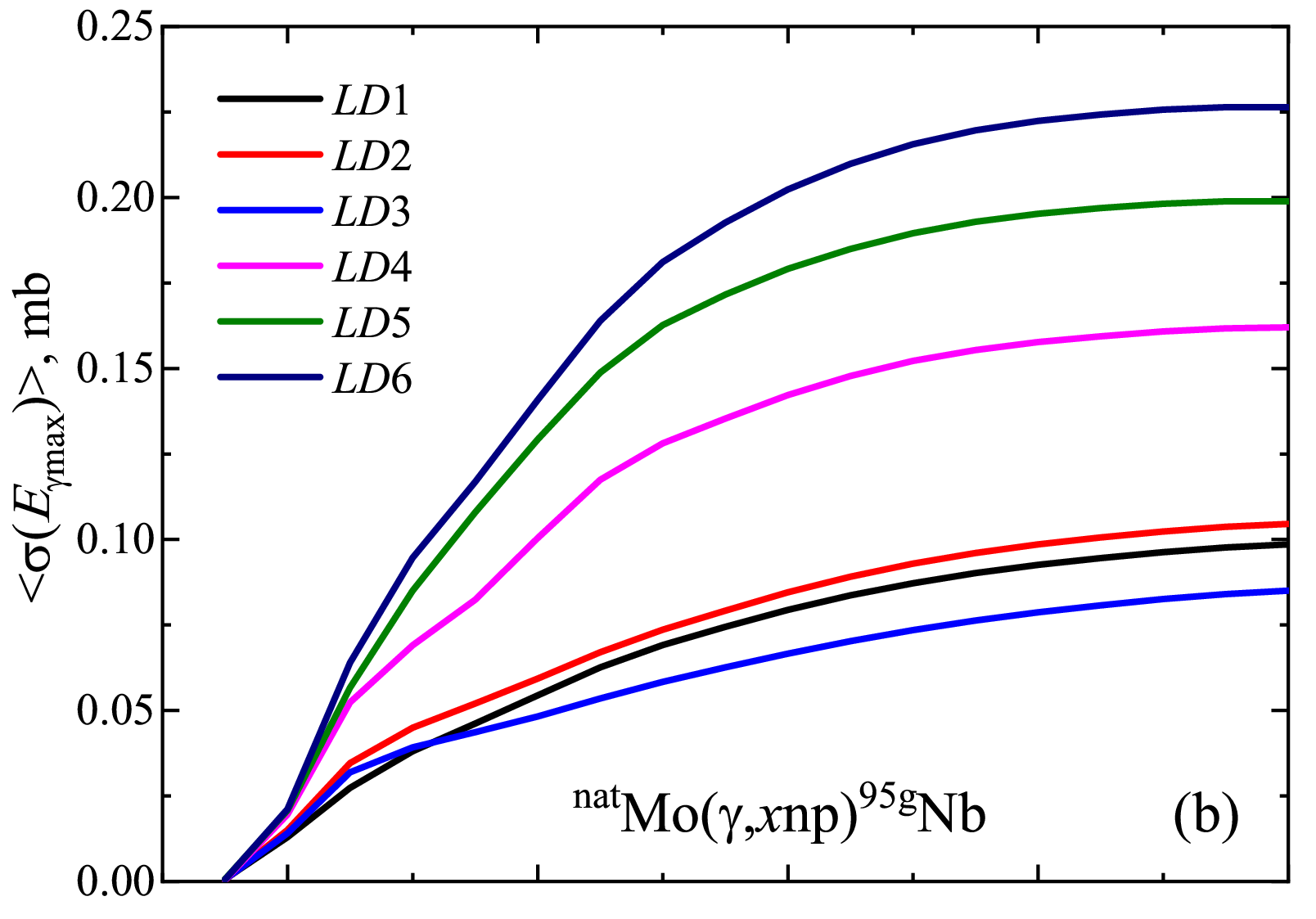}}
			\vspace{-1ex}
				\resizebox{0.45\textwidth}{!}{%
					\includegraphics{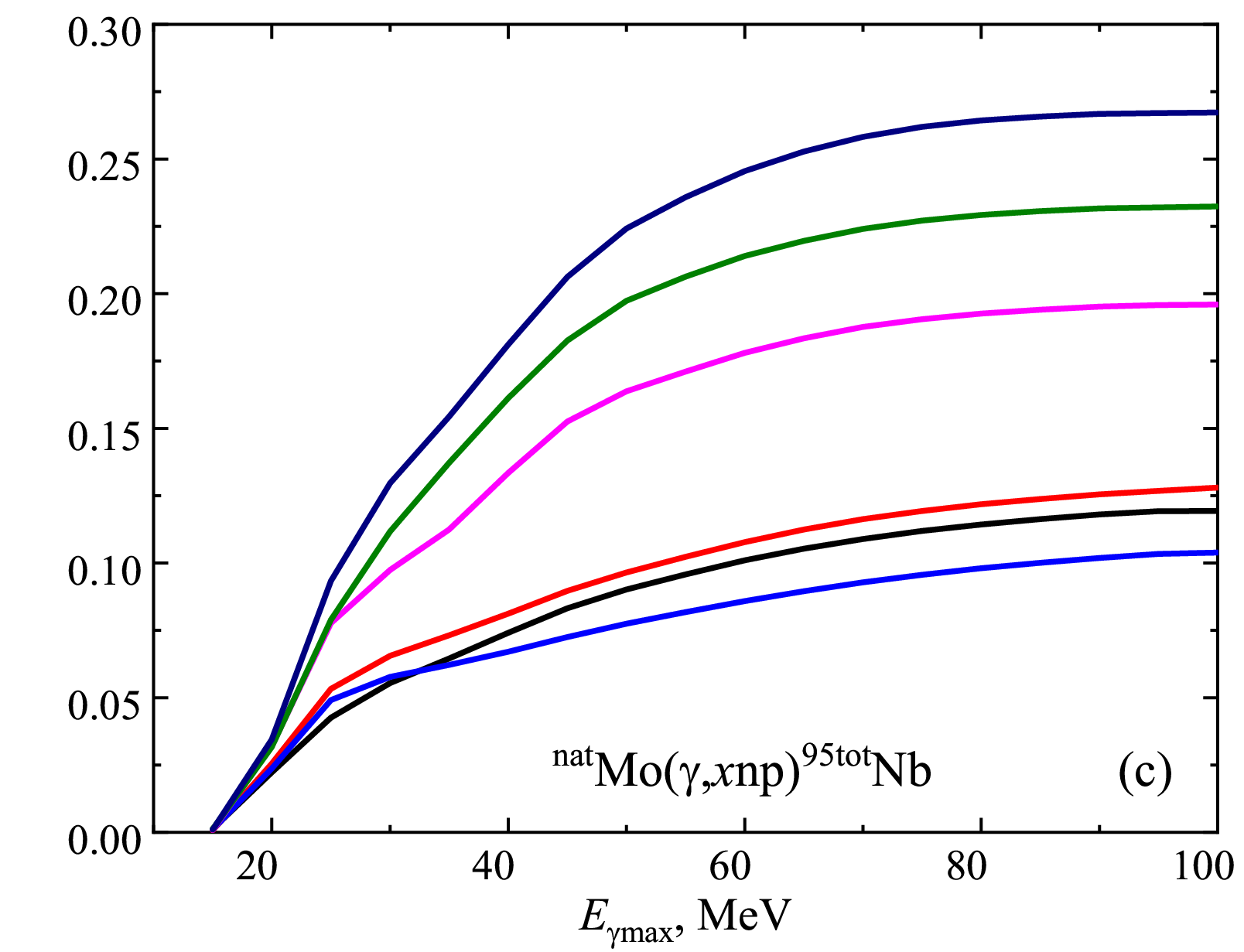}}
				\caption{Theoretical flux-averaged cross-sections $\langle{\sigma(E_{\rm{\gamma max}})}\rangle$ for the $^{\rm nat}$Mo$(\gamma,x\rm np)^{95\rm m,g,tot}$Nb reactions. The calculations were performed using the minimal reaction thresholds and TALYS1.95 code for six different level density models.}
				\label{fig6}
			\end{figure}								

The cross-sections $\sigma(E)$ are averaged over the bremsstrahlung flux $W(E,E_{\rm{\gamma max}})$ in the energy range from the threshold of the  reaction $E_{\rm{thr}}$ to the maximum energy of the bremsstrahlung spectrum $E_{\rm{\gamma max}}$. As a result, flux-averaged cross-sections were obtained: 

\begin{equation}\label{form3}
	\langle{\sigma(E_{\rm{\gamma max}})}\rangle_{\rm th} = 
	\frac{ \int\limits_{E_{\rm{thr}}}^{E_{\rm{\gamma max}}}\sigma(E) \cdot  W(E,E_{\rm{\gamma max}})dE}{{\int\limits_{E_{\rm{thr}}}^{E_{\rm{\gamma max}}} W(E,E_{\rm{\gamma max}})dE}.}
\end{equation}

For calculation of flux-averaged cross-sections 	$\langle{\sigma(E_{\rm{\gamma max}})}\rangle$ for the $^{\rm nat}$Mo$(\gamma,x\rm np)^{95\rm m,g}$Nb reactions, we average the cross-sections $\sigma(E)$
 using minimal reaction thresholds: in the case formation of $^{95\rm m}$Nb -- $E_{\rm thr}$ = 9.54~MeV and $^{95\rm g}$Nb -- $E_{\rm thr}$ =  9.30 MeV. This was done to be able to compare the theoretical  flux-averaged cross-sections with experimental results, where the total flux from the $^{96}$Mo threshold to $E_{\rm{\gamma max}}$ is also used. If, in making calculations, we use a self-own reaction threshold for each isotope, as we did in \cite{natMo-90Mo-90Nb,natNi-5756Ni}, then it is necessary to introduce an appropriate correction factor. Since such a coefficient is theoretically dependent, we do not use this approach in this work.
 
The results of the flux-averaged cross-section $\langle{\sigma(E_{\rm{\gamma max}})}\rangle$ calculation for the $^{\rm nat}$Mo$(\gamma,x\rm np)^{95\rm m,g,tot}$Nb reactions are presented in Fig.~\ref{fig6}. 

The shape of the energy dependence of the cross-sections for the formation of metastable and ground states is noticeably different. In the case of the metastable state, we observe the dominance of the contribution from $^{96}$Mo (Fig.~\ref{fig5} (a), energies of about 20~MeV). This leads to rapid saturation of the flux-averaged cross-section for the metastable state (Fig.~\ref{fig6}~(a)). In the case of ground and total cross-sections, the contributions of isotope cross-sections are distributed more evenly, and the flux-averaged cross-sections increase smoothly up to 95~MeV. Note that the cross-sections for the $^{96,97,98,100}$Mo nuclei are given taking into account the percentage abundance of isotopes. The cross-sections of the $^{\rm nat}$Mo$(\gamma,x\rm np)^{95\rm tot}$Nb reaction are their algebraic sum.

\section{Experimental results}
\label{sec:4}

\begin{figure}[t]
	\resizebox{0.46\textwidth}{!}{%
		\includegraphics{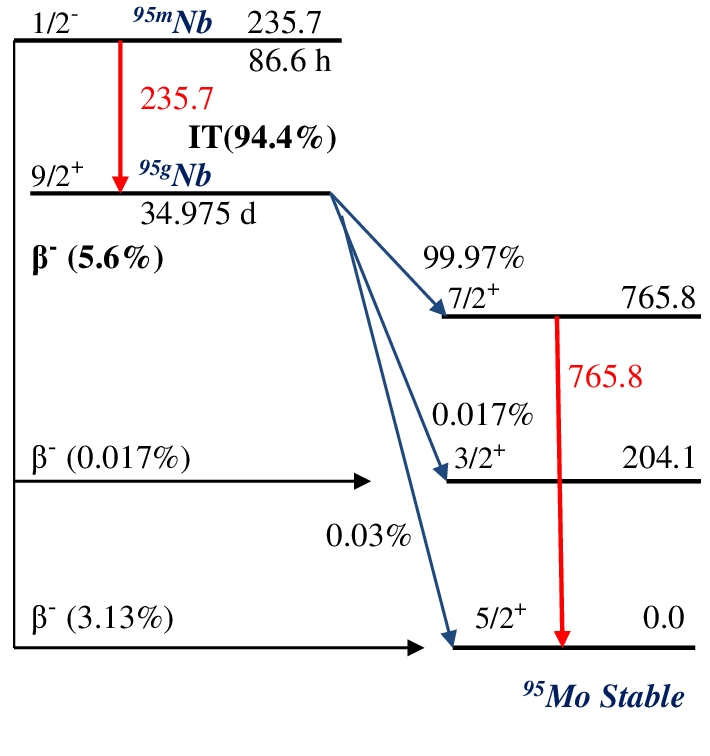}}
		\vspace{-5ex}
			\caption{Simplified representation of formation and decay scheme of the isomeric pair $^{95\rm m,g}$Nb. The nuclear level energies are in keV.}
	\label{fig7}

\end{figure}

A simplified diagram of the decay of the  $^{95}$Nb nucleus from the ground and metastable states is shown in Fig.~\ref{fig7}. Nuclear spectroscopic data of the radio-nuclides reactions $^{\rm nat}$Mo$(\gamma,x\rm np)^{95\rm m,g}$Nb are presented according to \cite{LUND}. 

The isomeric state $^{95\rm m}$Nb ($J^\pi = 1/2^{-}$) with a half-life $T_{1/2}$ of $86.6 \pm 0.08$ h decays to the unstable ground state $^{95\rm g}$Nb ($J^\pi = 9/2^{+}$) by emitting $\gamma$-quanta with the energy of 235.7~keV through an internal transition with a branching ratio $p$ of $94.4 \pm 0.6$\%. Meanwhile, 5.6\% of the isomeric state decays to the various energy levels of stable $^{95}$Mo via $\beta^-$-process. The unstable ground state $^{95\rm g}$Nb with a half-life $T_{1/2}$ of $34.975 \pm 0.007$~d decays to the 765.8 keV energy level of $^{95}$Mo via $\beta^-$-process (99.97\%). 

The cross-sections for the formation of the $^{95\rm m}$Nb nucleus in the metastable state in the reaction on $^{\rm nat}$Mo can be determined from direct measurements of the number of counts of $\gamma$-quanta $\triangle A$ in the full absorption peak at an energy of 235.7~keV ($I_{\gamma} = 24.9 \pm 0.8$\%). To calculate the experimental values $\langle{\sigma(E_{\rm{\gamma max}})}\rangle$ the following expression was used:

\begin{equation}
	\langle{\sigma(E_{\rm{\gamma max}})}\rangle = 
	\frac{\lambda \triangle A  {\rm{\Phi}}^{-1}(E_{\rm{\gamma max}})}{N_x I_{\gamma} \ \varepsilon (1-e^{-\lambda t_{\rm{irr}}})e^{-\lambda t_{\rm{cool}}}(1-e^{-\lambda t_{\rm{meas}}})},
	\label{form4}
\end{equation}
where ${\rm{\Phi}}(E_{\rm{\gamma max}}) = {\int\limits_{E_{\rm{thr}}}^{E_{\rm{\gamma max}}}W(E,E_{\rm{\gamma max}})dE}$  is the bremsstrahlung flux in the energy range from the reaction threshold $E_{\rm{thr}}$ up to $E_{\rm{\gamma max}}$; $N_x$ is the the number of studied atoms (including 4 isotopes of Mo -- 96, 97, 98, and 100); $I_{\gamma}$ is the intensity of the analyzed $\gamma$-quanta; $\varepsilon$ is the absolute detection efficiency for the analyzed $\gamma$-quanta energy; $\lambda$ denotes the decay constant \mbox{($\rm{ln}2/\textit{T}_{1/2}$)}; $T_{1/2}$ is the half-life of the nucleus;  $t_{\rm{irr}}$, $t_{\rm{cool}}$ and $t_{\rm{meas}}$ are the irradiation time, cooling time and measurement time, respectively.

In natural molybdenum targets, as a result of the $^{\rm nat}$Mo$(\gamma,x\rm n2p)$ reaction, the $^{95}$Zr nucleus can also be formed. In the decay scheme of $^{95}$Zr ($T_{1/2} = 65.02 \pm 0.05$~d) there is a
 $\gamma$-transition with the energy $E_\gamma$ = 235.7~keV and intensity $I_\gamma = 0.294 \pm 0.016$\%. The decay of the $^{95}$Zr nucleus can contribute to the observed value of $\Delta A\rm _m$. To take this
 contribution into account, calculations were performed in the TALYS1.95 code with level density model $LD$1.  It was found that the estimated activity of $^{95}$Zr by $\gamma$-line with 235.7 keV is negligible. The contribution of $^{95}$Zr was also experimentally verified by the $\gamma$-lines corresponding to the decay of the $^{95}$Zr nucleus, namely, $E_\gamma$ = 724.2 keV ($I_\gamma = 44.17 \pm 0.13$\%) and $E_\gamma$ = 756.7 keV ($I_\gamma$ = 54\%). No such peaks were found in the measured spectra of the induced $\gamma$-activity of the targets. 

For the $\gamma$-quanta with an energy of 235.7~keV and the thicknesses of the molybdenum targets used, the self-absorption coefficients were calculated using the GEANT4.9.2 code. It was found that the value of the self-absorption coefficient did not exceed 0.8\%, which was taken into account when processing the experimental results.

The uncertainty of measured flux-averaged cross-sections was determined as a square root of the quadratic sum of statistical and systematic errors. The statistical error in the observed $\gamma$-activity is mainly due to statistics in the full absorption peak of the corresponding $\gamma$-ray, which varies between 1.1 to 4.2\%. The measured $\triangle A$ value of the investigated $\gamma$ ray depends on the detection efficiency, half-life, and the intensity. The background is generally governed by the contribution from the Compton scattering of the emitted $\gamma$ rays.

Systematic errors are caused by the following uncertainties: exposure time and electron current $\sim$0.5\%, detection efficiency of  $\gamma$-rays by the detector 2--3\%, half-life $T_{1/2}$  of reaction products, and intensity $I_{\rm \gamma}$ of the analyzed  $\gamma$-rays. The error of experimental data normalization to the yield of the monitoring reaction $^{100}{\rm{Mo}}(\gamma,n)^{99}\rm{Mo}$ was 6\%.

\begin{figure}[h]
	\resizebox{0.47\textwidth}{!}{%
		\includegraphics{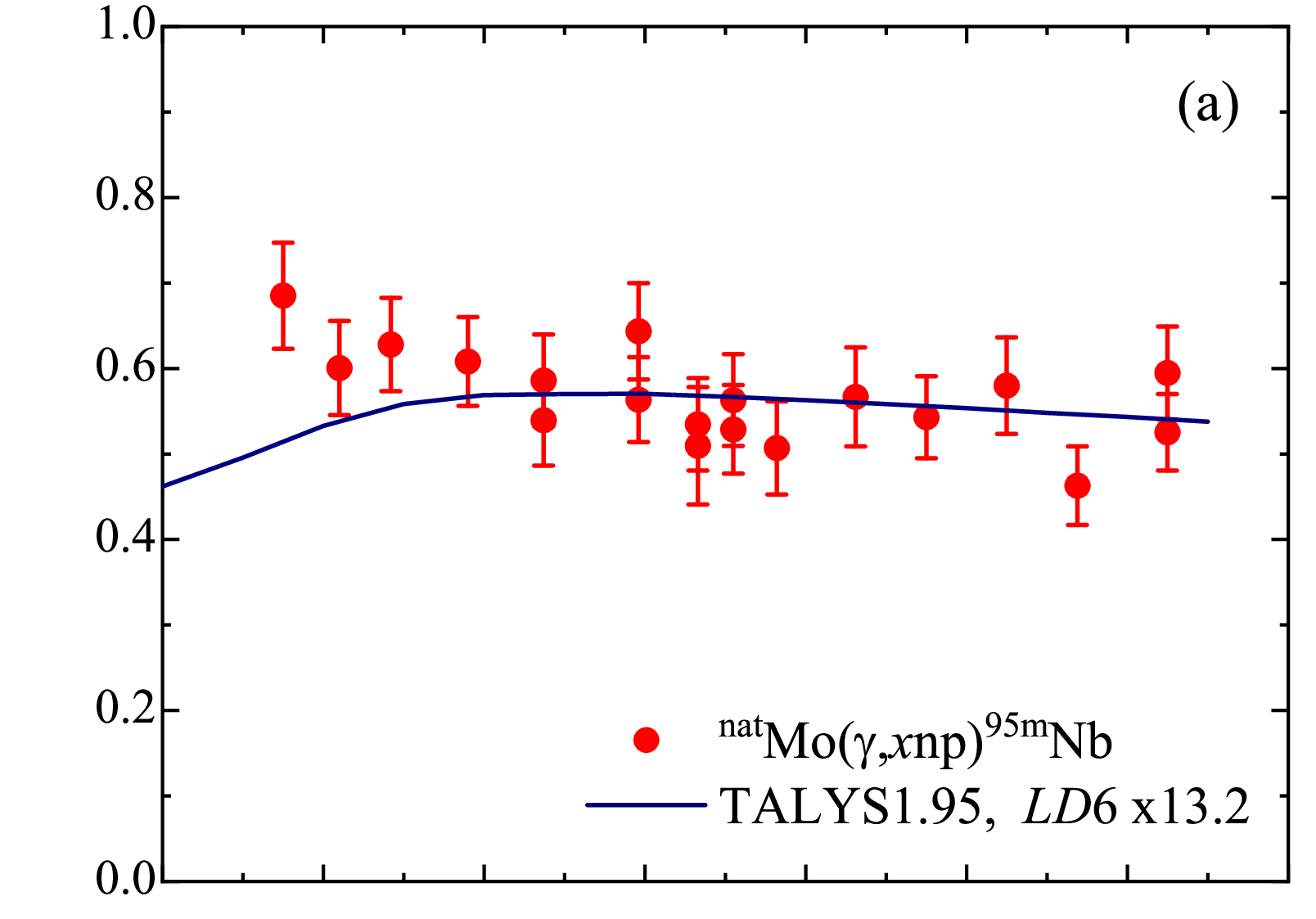}}
	\vspace{-1ex}
		\resizebox{0.47\textwidth}{!}{%
			\includegraphics{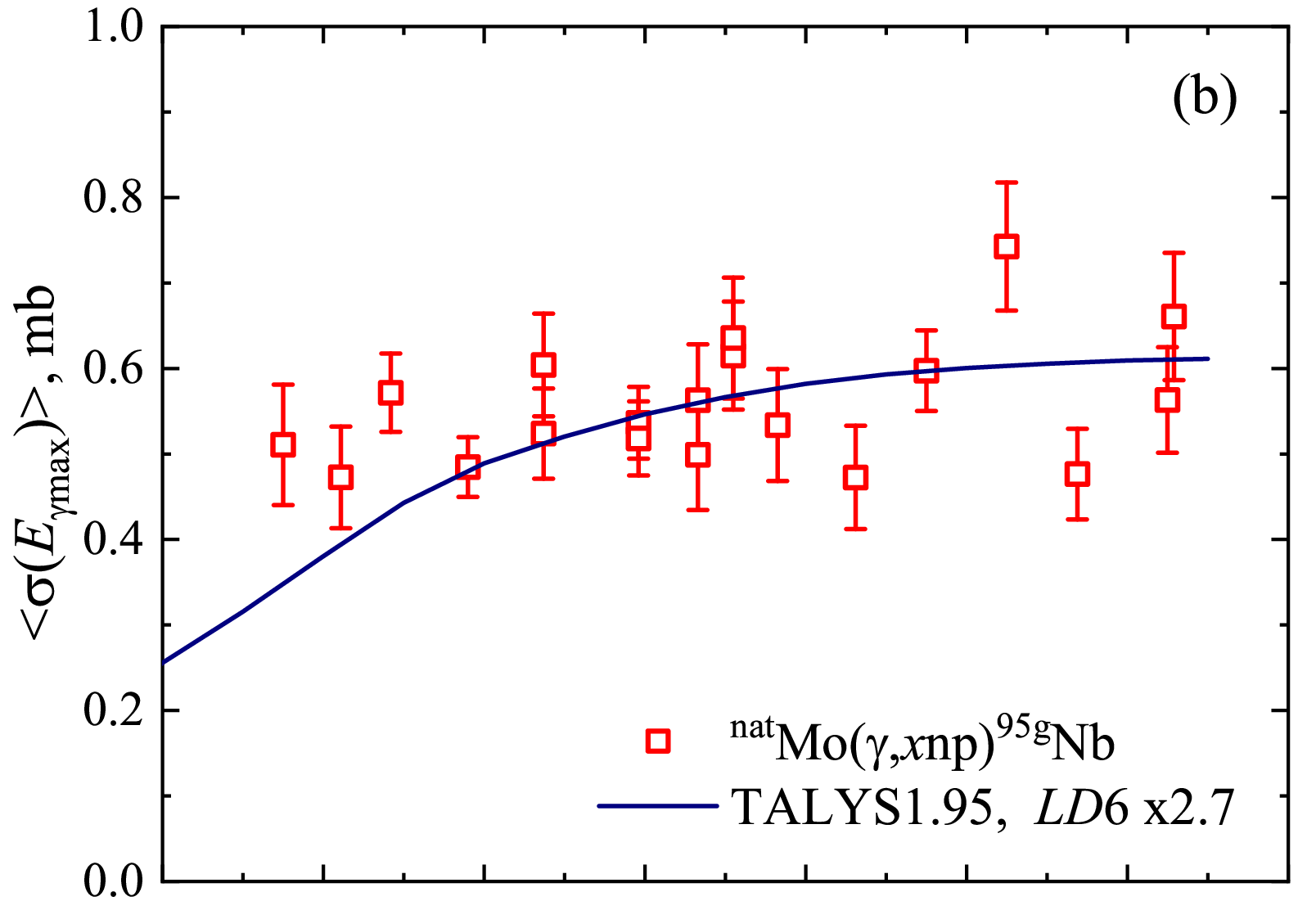}}
		\vspace{-1ex}
			\resizebox{0.47\textwidth}{!}{%
				\includegraphics{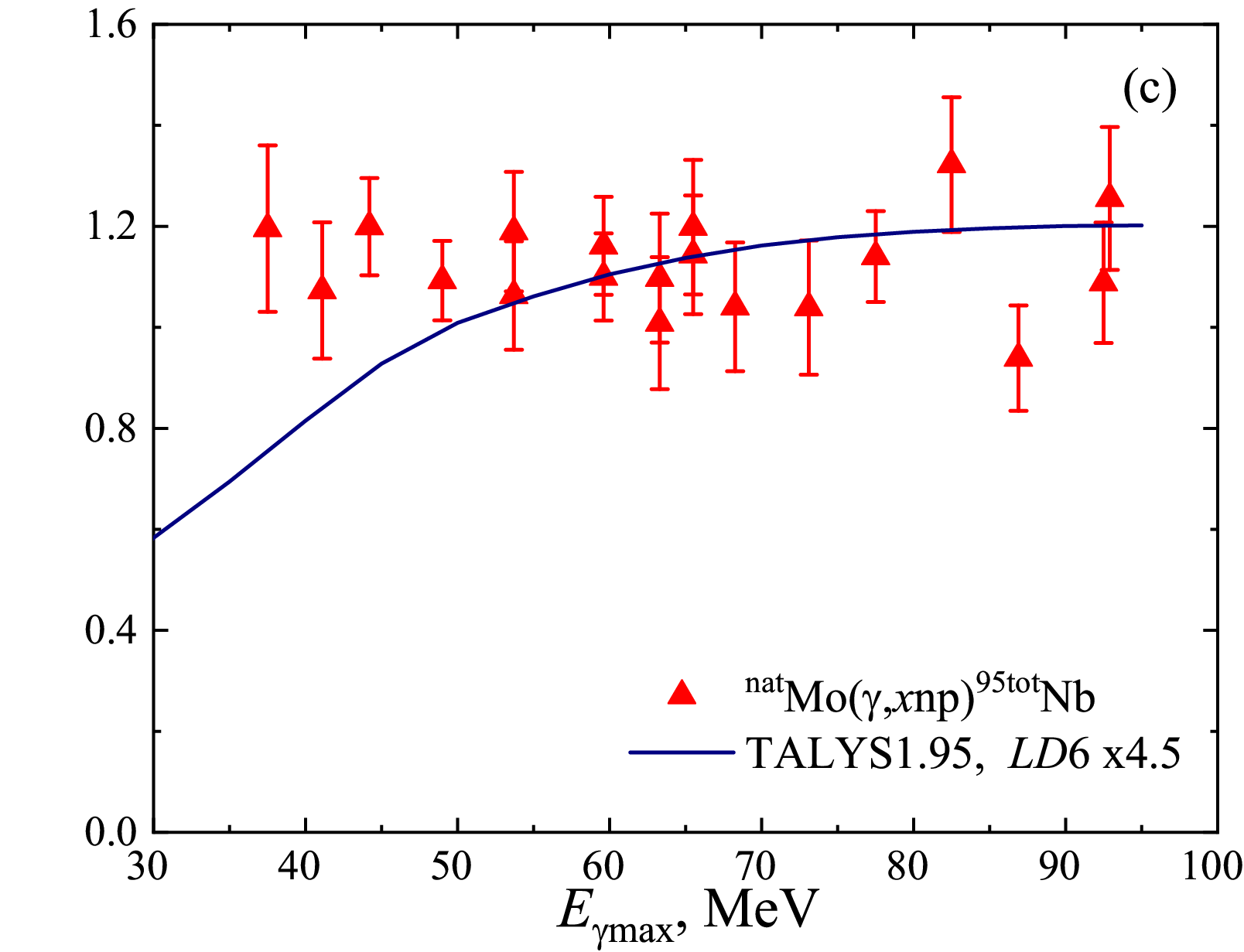}}
			\caption{Flux-averaged cross-sections $\langle{\sigma(E_{\rm{\gamma max}})}\rangle$ for the formation of $^{95\rm m}$Nb (a), $^{95\rm g}$Nb (b) on $^{\rm nat}$Mo, and total cross-section for the $^{\rm nat}$Mo$(\gamma,x\rm np)^{95\rm tot}$Nb reaction (c). Calculations were performed with the TALYS1.95 code for $LD$6.  }
			\label{fig8}   
		\end{figure}

The total uncertainties of the measured flux-averaged cross-sections are given in Fig.~\ref{fig8} and Table~\ref{tab1}.

    The experimental values of the flux-averaged cross-section $\langle{\sigma(E_{\rm{\gamma max}})}\rangle_{\rm m}$ of the $^{\rm nat}$Mo$(\gamma,x\rm np)^{95\rm m}$Nb reaction were determined at the bremsstrahlung end-point energy of 38--93 MeV (see Fig.~\ref{fig8} (a) and
     Table~\ref{tab1}). When calculating $\langle{\sigma(E_{\rm{\gamma max}})}\rangle_{\rm m}$ values, we use flux of bremsstrahlung quanta with a threshold for the reaction on isotope $^{96}$Mo, $E_{\rm thr}$ = 9.54~MeV. 
    
    As can be seen from the decay diagram of the isomeric and ground states of the $^{95}$Nb nucleus (see Fig.~\ref{fig7}), it is impossible to determine the value of the flux-averaged cross-section for the formation of the $^{95\rm g}$Nb nucleus in the ground state using direct measurements of the value $\triangle A$ in the full absorption peak at an energy of 765.8~keV. However, these values can be estimated using data on  $\langle{\sigma(E_{\rm{\gamma max}})}\rangle_{\rm m}$ for the formation of $^{95\rm m}$Nb  in a metastable state, obtained in this work, and the values of isomeric yield ratio $IR = Y_{\rm H}(E_{\rm{\gamma max}})/Y_{\rm L}(E_{\rm{\gamma max}})$, from the work \cite{arxivNb}. The $^{95\rm m}$Nb nucleus in the metastable state has a low-spin state and in the ground state $^{95\rm g}$Nb has a high-spin state, then knowing the $IR$ values, one can find the flux-averaged cross-section for the formation of the $^{95\rm g}$Nb nucleus $\langle{\sigma(E_{\rm{\gamma max}})}\rangle_{\rm g}$ from a simple relation: $\langle{\sigma(E_{\rm{\gamma max}})}\rangle_{\rm g} = \langle{\sigma(E_{\rm{\gamma max}})}\rangle_{\rm m} \cdot IR$. Similarly, one can estimate the total flux-averaged cross-section for the formation of a $^{95}$Nb nucleus on natural Mo.     
    
    The values of $\langle{\sigma(E_{\rm{\gamma max}})}\rangle_{\rm g}$ and $\langle{\sigma(E_{\rm{\gamma max}})}\rangle_{\rm tot}$, obtained by recalculation using $IR$, are presented in Fig.~\ref{fig8} (b, c).  
    
A comparison of experimental flux-averaged cross sections for $^{95\rm m,g,tot}$Nb shows a strong discrepancy with theoretical calculations. The closest to the experiment is the calculated result obtained using the cross-sections from the TALYS1.95 code for level density model $LD$6. The difference was 13.2 for the case of he formation  the  $^{95\rm m}$Nb nucleus in a metastable state. This factor was obtained by the least squares method.
An equally strong discrepancy for the cross-section for the metastable state was also observed in the case of studying the reaction $^{181}$Ta$(\gamma,\rm p)^{180\rm m}$Hf \cite{arxivHf}. In the case of the formation of a $^{95\rm g}$Nb nucleus in the ground state and for the total cross-section, the calculation results are much closer to the experimental data. The comparison shows that the difference between theory and experiment was 2.7 and 4.5 times, respectively.

As noted earlier according to \cite{exfor}, a study of the reaction $^{96}$Mo$(\gamma,\rm p)^{95}$Nb  was performed in \cite{Ma78}, and experimental values of the average cross-section per equivalent photon $\langle{\sigma(E_{\rm{\gamma max}})}\rangle_{\rm Q}$ were obtained. For comparison with these data, we presented our experimental results as $\langle{\sigma(E_{\rm{\gamma max}})}\rangle_{\rm Q}$. To determine the value of the averaged cross-sections per equivalent photon, a formula was used, that written in \cite{Ma78} as:

\begin{equation}
	\langle{\sigma(E_{\rm{\gamma max}})}\rangle_{\rm Q} = 
\frac {\int\limits_{E_{\rm{thr}}}^{E_{\rm{\gamma max}}}\sigma(E)\cdot W(E,E_{\rm{\gamma max}})dE}
{\frac{1}{E_{\rm{\gamma max}}} \int\limits_{E_{\rm{thr}}}^{E_{\rm{\gamma max}}} E \cdot W(E,E_{\rm{\gamma max}})dE}.
	\label{form5}
\end{equation}

First, we analyzed the cross-section for the $^{100}$Mo$(\gamma,\rm n)^{99}$Mo reaction, which is used in our experiment as a monitoring reaction.  
In the case of working with natural Mo targets, this reaction occurs only on one isotope of molybdenum with an atomic mass of 100. The comparison of the results presented as $\langle{\sigma(E_{\rm{\gamma max}})}\rangle_{\rm Q}$ with data from \cite{Ma78} agreed within 4.5--7.1\%, which does not exceed the experimental error of our data. This makes it possible to compare cross-sections for the reaction $^{\rm nat}$Mo$(\gamma,x\rm np)^{95m}$Nb under study. Note that the authors of \cite{Ma78} did not provide experimental errors for their results.

\begin{figure}[t]
	\resizebox{0.48\textwidth}{!}{%
		\includegraphics{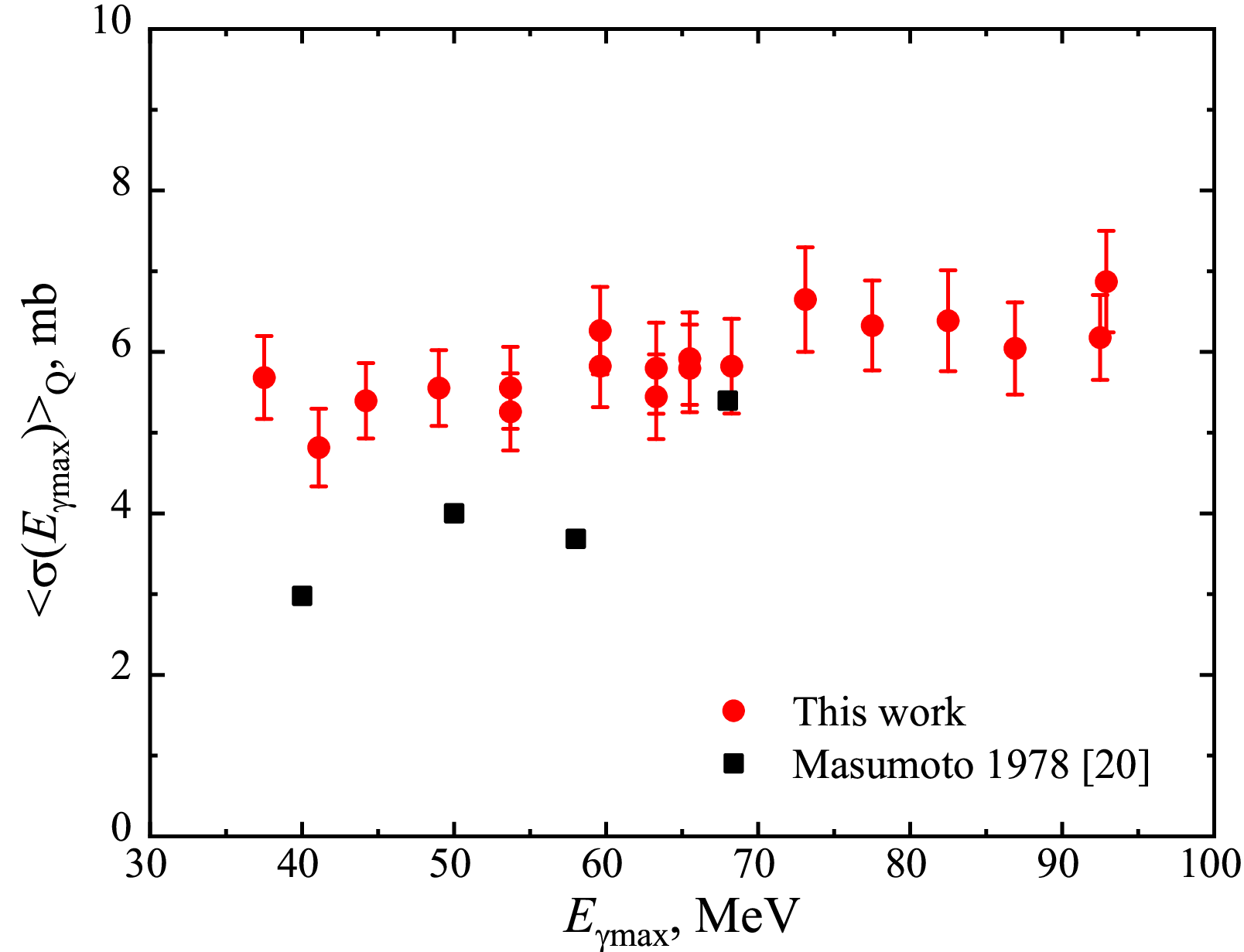}}
	\caption{Average cross-section per equivalent photon $\langle{\sigma(E_{\rm{\gamma max}})}\rangle_{\rm Q}$  for the ${^{96}\rm{Mo}}(\gamma,\rm p)^{95\rm m}\rm{Nb}$ reaction. The results of this work are denoted by red circles, data from \cite{Ma78} -- black squares.}
	\label{fig9}
\end{figure}

In \cite{Ma78} used targets made of natural molybdenum. The cross-sections $\langle{\sigma(E_{\rm{\gamma max}})}\rangle_{\rm Q}$ for the $^{96}$Mo$(\gamma,\rm p)^{95\rm m}$Nb reaction were define on the assumption that the main channel in the formation of the $^{95\rm m}$Nb nucleus on natural molybdenum is the reaction on $^{96}$Mo, the isotopic abundance of which is 16.68\%. Accordingly, when assessing $\langle{\sigma(E_{\rm{\gamma max}})}\rangle_{\rm Q}$, we obtained our results taking into account this isotopic concentration in order to correctly compare measurement data from different laboratories. As can be seen from Fig.~\ref{fig9}, our result $\langle{\sigma(E_{\rm{\gamma max}})}\rangle_{\rm Q}$ and the data of \cite{Ma78} are in reasonable agreement with each other.

It should be noted that according to the results of calculations of the normalized reaction yield $Y_{\rm i}(E_{\rm{\gamma max}})$, performed using cross-sections from the TALYS1.95 code (see Fig.~\ref{fig4}), the statement about the dominant role of the $^{96}$Mo$(\gamma,\rm p)$ reaction in the case of the formation of the $^{95\rm m}$Nb nucleus on natural Mo is valid at energies up to 30~MeV (provided the contribution of the dominant reaction is at least 90\%).

 \begin{table}[h]
	\caption{\label{tab1} Flux-averaged cross-sections for the ${^{\rm nat}\rm{Mo}}(\gamma,x\rm np)^{95\rm m,g,tot}\rm{Nb}$ reactions.}
	\centering
	\begin{ruledtabular}
		\begin{tabular}{lccr}
			\noalign{\smallskip}
			& & $\langle{\sigma(E_{\rm{\gamma max}})}\rangle$,  mb &  \\ \hline 	\noalign{\smallskip}
			$E_{\rm{\gamma max}}$,  MeV &  metastable & ground  & total \\   \hline\hline 	\noalign{\smallskip}
			37.50 & 0.68 $\pm$ 0.06 &  0.51 $\pm$ 0.07 &  1.20 $\pm$ 0.15 \\ 
			41.10 & 0.60 $\pm$ 0.06 &  0.47 $\pm$ 0.06 &  1.07 $\pm$ 0.13 \\ 
			44.20 & 0.63 $\pm$ 0.05 &  0.57 $\pm$ 0.05 &  1.20 $\pm$ 0.10 \\ 
			49.00 & 0.61 $\pm$ 0.05 & 0.48 $\pm$ 0.05 &  1.09 $\pm$ 0.11 \\ 
			53.70 &  0.54 $\pm$ 0.05 &  0.52 $\pm$ 0.05 &  1.06 $\pm$ 0.11 \\ 
			53.70  & 0.59 $\pm$ 0.05 &  0.60 $\pm$ 0.06 &  1.19 $\pm$ 0.12 \\ 
			59.60 & 0.56 $\pm$ 0.05 &  0.54 $\pm$ 0.05 &  1.10 $\pm$ 0.10 \\ 
			59.60  & 0.64 $\pm$ 0.06 &  0.52 $\pm$ 0.05 &  1.16 $\pm$ 0.10 \\ 
			63.30 & 0.53 $\pm$ 0.05 &  0.56 $\pm$ 0.07 &  1.10 $\pm$ 0.13 \\ 
			63.30  & 0.51 $\pm$ 0.07 &  0.50 $\pm$ 0.07 &  1.00 $\pm$ 0.14 \\ 
			65.50 & 0.53 $\pm$ 0.05 &  0.61 $\pm$ 0.06 &  1.14 $\pm$ 0.12 \\ 
			65.50  & 0.56 $\pm$ 0.05 &  0.64 $\pm$ 0.07 & 1.20 $\pm$ 0.13 \\ 
			68.25 &  0.51 $\pm$ 0.05 &  0.53 $\pm$ 0.07 &  1.04 $\pm$ 0.13 \\ 
			73.10 &  0.57 $\pm$ 0.06 &  0.47 $\pm$ 0.06 &  1.04 $\pm$ 0.13  \\ 
			77.50 &  0.54 $\pm$ 0.05 &  0.60 $\pm$ 0.05 &  1.14 $\pm$ 0.10  \\ 
			82.50 &  0.58 $\pm$ 0.06 &  0.74 $\pm$ 0.07 &  1.32 $\pm$ 0.13  \\ 
			86.90 &  0.46 $\pm$ 0.05 &  0.48 $\pm$ 0.05 &  0.94 $\pm$ 0.10   \\ 
			92.50 &  0.53 $\pm$ 0.04 &  0.56 $\pm$ 0.06 &  1.09 $\pm$ 0.12   \\ 
			92.90 &  0.59 $\pm$ 0.05 &  0.66 $\pm$ 0.07 &  1.26 $\pm$ 0.14   \\ \noalign{\smallskip}			\end{tabular} \\ 
	\end{ruledtabular}
\end{table}
 
\section{Conclusions}
\label{Concl}

In the present work, the experiment was performed on the beam of a linear electron accelerator LUE-40 RDC "Accelerator" NSC KIPT using the activation and off-line $\gamma$-ray spectrometric techniques. The bremsstrahlung flux-averaged cross-section $\langle{\sigma(E_{\rm{\gamma max}})}\rangle_{\rm m}$ of the $^{95\rm m}$Nb production in photonuclear reactions on natural Mo targets was determined for the first time. The bremsstrahlung end-point energy range was $E_{\rm{\gamma max}}$ = 38--93 MeV.

Based on the experimental data on $\langle{\sigma(E_{\rm{\gamma max}})}\rangle_{\rm m}$ and the isomeric yield ratio $IR$, the values of $\langle{\sigma(E_{\rm{\gamma max}})}\rangle_{\rm g}$ and $\langle{\sigma(E_{\rm{\gamma max}})}\rangle_{\rm tot}$ for the formation of the $^{95}$Nb nucleus in the ground state and for the total cross-section of the ${^{\rm nat}\rm{Mo}}(\gamma,x\rm np)$ reaction were obtained in this work.

  The calculation of the flux-averaged cross sections $\langle{\sigma(E_{\rm{\gamma max}})}\rangle$ was performed using the cross sections $\sigma(E)$ for the studied reactions from the TALYS1.95 code for different level density models $LD$ 1--6. 
  
In the case of the ${^{\rm nat}\rm{Mo}}(\gamma,x\rm np)^{95\rm m}$Nb reaction, the comparison showed a significant excess (about 13.2~times) of the experimental results over the theoretical flux-averaged cross sections $\langle{\sigma(E_{\rm{\gamma max}})}\rangle_{\rm m}$. This result can be used to improve the TALYS code.

The calculation result with the level density model $LD$6 (microscopic level densities based on temperature-dependent Hartree-Fock-Bogoliubov calculations using the Gogny force from Hilaire's combinatorial tables) is the closest to the experimental cross-section data.


\section*{Acknowlegment}
The authors would like to thank the staff of the linear electron accelerator LUE-40 NSC KIPT, Kharkiv, Ukraine, for their cooperation in the realization of the experiment.

\section*{Declaration of competing interest}
The authors declare that they have no known competing financial interests or personal relationships that could have appeared to influence the work reported in this paper.

\end{document}